\documentclass[12pt,preprint]{aastex}

\slugcomment{draft version \today, }

\shorttitle{Jet propagations, breakouts and photospheric emissions in collapsing massive progenitors of long duration gamma ray bursts}
\shortauthors{Nagakura et al.}

\begin{document}

\title{Jet propagations, breakouts and photospheric emissions in collapsing massive progenitors of long duration gamma ray bursts}

\author{Hiroki Nagakura\altaffilmark{1}, Hirotaka Ito\altaffilmark{3}, Kenta Kiuchi\altaffilmark{4}, Shoichi Yamada\altaffilmark{1,2}} 

\altaffiltext{1}{Department of Science and Engineering,
 Waseda University, 3-4-1 Okubo, Shinjuku, Tokyo 169-8555, Japan}
\altaffiltext{2}{Advanced Research Institute for Science and Engineering, 
Waseda University, 3-4-1 Okubo, Shinjuku, Tokyo 169-8555, Japan}
\email{hiroki@heap.phys.waseda.ac.jp}
\altaffiltext{3}{Department of Aerospace Engineering, Tohoku University, 6-6-01 Aramaki-Aza-Aoba, Aoba-ku, Sendai, 980-8579, Japan}
\altaffiltext{4}{Yukawa Institute for Theoretical Physics, Kyoto University, Kyoto, 606-8502, Japan}

\begin{abstract}
 We investigate by two-dimensional axisymmetric relativistic hydrodynamical simulations (1) jet propagations 
through an envelope of a rapidly rotating and collapsing massive star, which is supposed to be a progenitor of long duration 
gamma ray bursts (GRBs), (2) breakouts and subsequent expansions into stellar winds and (3) accompanying 
photospheric emissions. We find that if the envelope rotates uniformly almost 
at the mass shedding limit, its outer part stops contracting eventually when the centrifugal force becomes 
large enough. Then another shock wave is formed, propagates outwards and breaks out of the envelope into the 
stellar wind. Which 
breaks out earlier, the jet or the centrifugal bounce-induced shock, depends on the timing of jet injection. 
If the shock breakout occurs earlier owing to a later injection, the jet propagation and subsequent photospheric 
emissions are affected substantially. We pay particular 
attention to observational consequences of the difference in the timing of jet injection.
We calculate optical depths to find the location of photospheres, extracting densities and temperatures at 
appropriate retarded times from the hydrodynamical data. We show that the luminosity and observed 
temperature of the photospheric emissions are both much lower than those reported in previous studies. 
Although luminosities are still high enough for GRBs, the observed temperature are lower than the energy at the 
spectral peak expected by the Yonetoku-relation. This may imply that energy exchanges between photons and 
matter are terminated deeper inside or some non-thermal processes are operating to boost photon energies.
\end{abstract}


\section{Introduction}
There is mounting observational evidence that links GRBs to the death of massive stars \citep{2006ARA&A..44..507W} 
and it is widely believed that GRBs are associated with the formation of black hole or magnetar via the collapse 
of rapidly rotating massive stars \citep{1993ApJ...405..273W,1998ApJ...494L..45P,1999ApJ...524..262M}. Although we 
do not know exactly how a large amount of energy is generated, the most promising scenario is that a relativistic 
jet is launched from the central engine by neutrino annihilation or magnetohydrodynamical processes, propagates 
through a progenitor star and stellar wind \citep{1999ApJ...524..262M}, and then dissipates its kinetic energy by 
internal shocks or photospheric emissions or relativistic turbulences \citep
{2004RvMP...76.1143P,2008ApJ...682..463P,2009ApJ...700L..47L,2009ApJ...695L..10L,2009MNRAS.395..472K}, producing 
the prompt emissions of GRBs or XRFs.

A large number of numerical works have been devoted so far to the understanding of relativistic jet propagations 
in the stellar envelope \citep
{2000ApJ...531L.119A,2003ApJ...586..356Z,2004ApJ...608..365Z,2006ApJ...651..960M,2009ApJ...699.1261M,2007ApJ...665..569M,2007ApJ...657L..77T,2009ApJ...700L..47L,2010arXiv1006.2440M}. These simulations have demonstrated that the 
jet is confined by the pressure of hot cocoon as it penetrates through the stellar envelope. The Kelvin-Helmholtz 
instability, which occurs between the cocoon and the jet, produces rich internal structures \citep
{2005ApJ...629..903L,2007ApJ...665..569M}. More recently, \citet{2009ApJ...700L..47L,2010arXiv1006.2440M} computed 
the jet propagation beyond the stellar surface and observed that these internal structures in the jet and cocoon 
leave their traces until later times. They also claimed that the hot jet produces very bright and highly efficient 
photospheric emissions in the prompt phase of GRBs. These very efficient photospheric emissions may solve the 
efficiency problem of the prompt emission \citep{2007ApJ...670L..77I}. Interestingly, thermal emissions were 
indeed identified for some Long GRBs lately \citep{2009ApJ...706L.138A,2010ApJ...709L.172R,2010arXiv1010.4601G}. 
The photospheric emissions from the relativistic jet are hence attracting much attention of the GRB society these 
days \citep{2008ApJ...682..463P,2010arXiv1002.2634T,2010PThPh.124..667I}.

It should be noted that the previous numerical studies on the jet propagation ignored the infall of the stellar 
envelope. According to the collapsar model, on which this paper is based, the gravitational core collapse sets in 
just like ordinary supernovae when the density reaches $\rho \sim 10^{9.5} {\rm g/cm^3}$ or the 
temperature exceeds $T \sim 5 \times 10^{9} {\rm K}$ and electron captures or endothermic 
photodissociations of nuclei reduce pressure \citep[see e.g.][]{2006RPPh...69..971K}. A shock wave produced by core 
bounce stalls in the core and a large amount of matter accretes on a time scale of seconds onto a proto-neutron 
star at first and into a black hole later. The so-called prompt shock wave either remains stagnated near the black 
hole or is swallowed into it. On the other hand, the core collapse produces a rarefaction wave at the boundary of 
the core and envelope, which then propagates outward through the envelope and induces the infall of the envelope 
when it arrives. Thus, the neglect of the envelope motion in studying the jet propagation is justified only when 
the jet is launched very early on, possibly soon after the black hole is formed 
\citep{2001ApJ...550..410M,2003ApJ...586..356Z}, and the infall of the envelope is not yet substantial. If the jet 
launch is delayed somehow, on the other hand, the profile of the envelope will be modified and the jet propagation 
will be affected. It is also pointed out that the stellar envelope may cease to infall eventually. In fact, the 
outer portion of the stellar envelope is likely to have an angular momentum large enough to terminate the infall 
by centrifugal forces \citep{2006ApJ...637..914W}. Indeed, \citet{2010ApJ...713..800L} observed in their long term 
simulations of rotational collapse of massive stars that a shock wave is generated by centrifugal forces and the 
outer portion of stellar envelope is expelled eventually. Since we do not know exactly when the jet is launched, 
it is important to study the influence of envelope dynamics on the jet propagation and subsequent prompt emissions.

 Motivated by these facts, we numerically investigate the relativistic jet propagations through a non-stationary 
envelope, moving either inward or outward, of a rapidly rotating massive star, varying the timing of jet 
injection. We assume that the prompt shock wave of core bounce origin has already been swallowed into 
the black hole and what is supposed to occur in the core after bounce such as neutrino heating and various 
hydrodynamical instabilities do not affect the dynamics of the envelope. We do not specify the mechanism of the 
jet launch from the central engine, which is under controversy at present, and inject the jet with appropriate 
properties from the computational inner boundary by hand, following the common practice in this field. Our focus 
is the jet propagations in the non-stationary envelope and its influences on the subsequent photospheric 
emissions. This paper is organized as follows. In Section 2, we describe the models and numerical procedures. Then 
our main results will be presented in Section 3. We conclude the paper with the summary of our findings in 
Section~4.


\section{Methods}
 As mentioned above, in this letter we compute the jet propagations through the envelope of a rotating massive 
star into a stellar wind, taking into account the core-collapse-induced motions of the envelope under the
assumption that the prompt shock wave is soon sucked into the black hole and various processes in the core 
such as neutrino heating of accreting matter and hydrodynamical instabilities do not affect the dynamics of 
the envelope. In order to simulate the infall of the envelope induced by the rarefaction wave that is generated by
core collapse, we take rather involved multiple steps. More specifically, we (1) construct massive star's envelope 
models in rotational equilibrium, (2) put quasi-steady winds on top of them, (3) simulate rotational collapse of 
the envelope, generating a rarefaction wave by artificially reducing the pressure gradient at the inner boundary 
and (4) compute subsequent jet propagations in the envelope and (5) calculate photospheric emissions as a 
postprocess. We employ the so-called HSCF scheme in the first step and perform 2D relativistic hydrodynamical 
simulations in the third and fourth steps. In the following, we explain what is done at each step more in detail 
in order to facilitate reader's understanding of our results in the next section.

\subsection{A Massive Star Envelope in Rotational Equilibrium}\label{roteq}
The first step is a preparation of the initial model for dynamical simulations in the later phases. In this 
subsection, we construct a 2D axisymmetric model of a rotating massive star envelope in dynamical equilibrium. We 
employ the method developed by \citet{1986ApJS...61..479H,2010ApJ...717..666K}. It should be noted here that the 
currently most elaborate stellar evolution models are still unable to fully implement rotational equilibrium and 
neglect the non-spherical deformation of rotating stars. In this study, however, the rotational equilibrium is 
crucial, since the infall of the envelope commences only after the rarefaction wave generated at the boundary of
the core and envelope arrives. If the initial model is not in dynamical equilibrium, however, even outer parts 
of the envelope begin to move immediately after the simulation is started and false shock waves are produced as 
a consequence more often than not. 

\begin{figure}
\vspace{15mm}
\epsscale{1.0}
\plotone{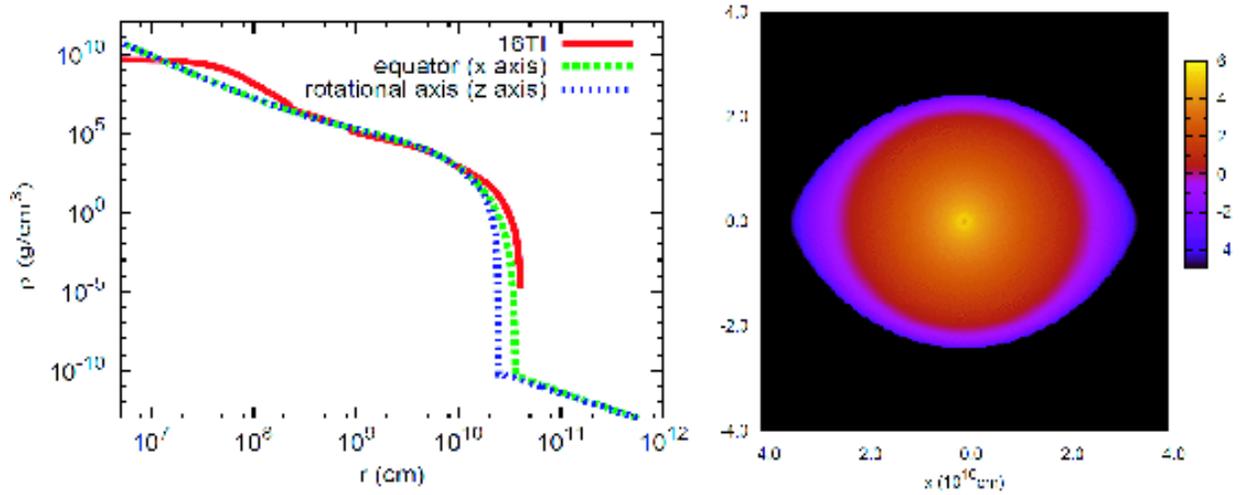}
\caption{ The density profiles on the rotational axis (z-axis) and equator (x-axis) for the envelope model in this paper and model 16TI  
by \citet{2006ApJ...637..914W} (left panel) and the density contour (log scale) in the meridian section of the same envelope model (right 
panel).
\label{f1}}
\end{figure}

\begin{figure}
\vspace{15mm}
\epsscale{1.0}
\plotone{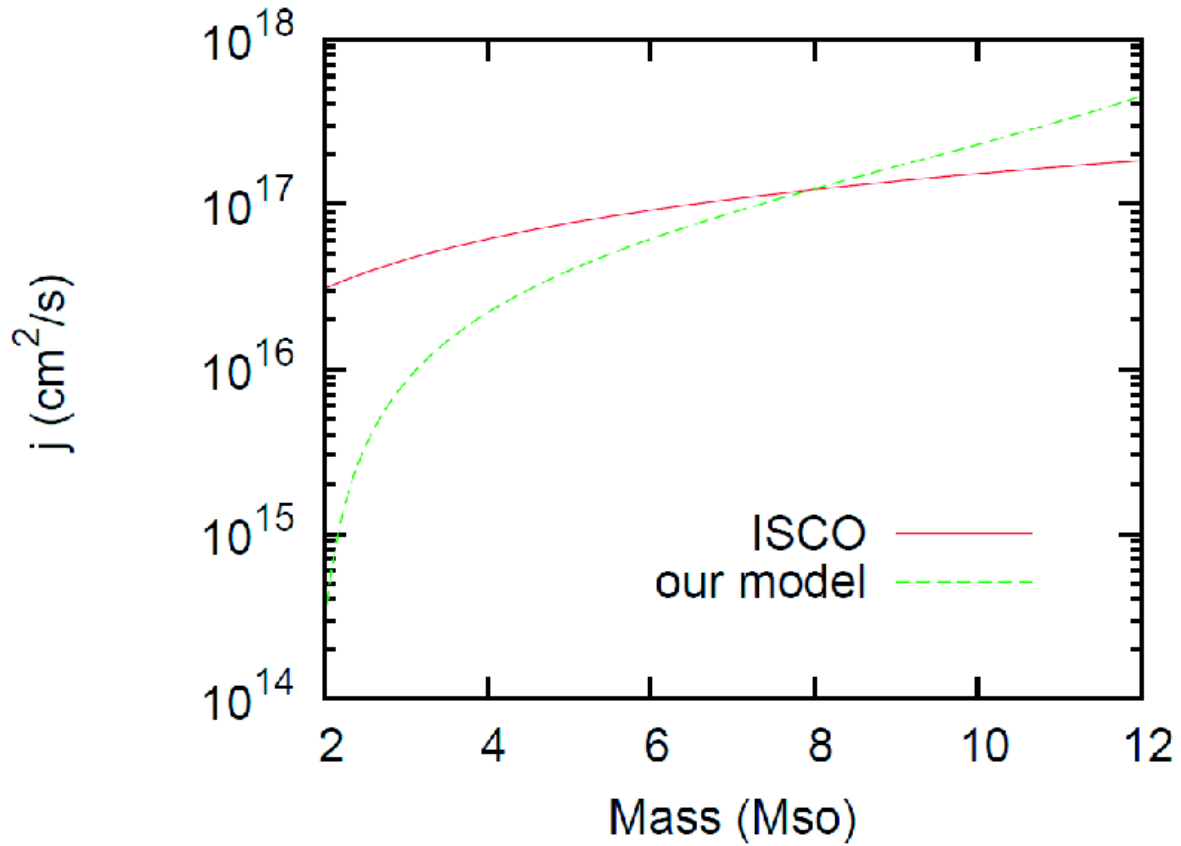}
\caption{The profile of specific angular momentum on the equatorial plane for the envelope model in this paper as a function of 
included mass (green line). The red line shows the specific angular momenta at the innermost stable circular orbit (ISCO) for  Schwartzschild Black Holes as a function of their masses.
\label{f2}}
\end{figure}

Our envelope model is constructed so as to mimic 16TI model by \citet{2006ApJ...637..914W}, which is currently 
supposed to be one of the most promising GRB progenitor models. Since the outer envelope of 16TI is almost radiation 
dominated, we employ a polytropic equation of state (EOS) with the adiabatic index of $ \gamma = 4/3$. We impose a 
rigid rotation as an approximation to the outer envelope of 16TI. Figure~\ref{f1} shows the density profiles on the 
rotational axis and equator for our model together with the one for 16TI. Also displayed in the figure is the 
density distribution in the meridian section for our model. Our model agrees fairly well with 16TI except for the 
innermost portion, where the rotation is not rigid in 16TI. This discrepancy is not very important for the 
investigation in this study, since that part is sucked into the inner boundary much earlier than the jet 
injection. Our envelope model has a total mass of $M \simeq 14M_{\sun}$ and a specific angular momentum 
of $j_{sp} \simeq 1.5 \times 10^{19} {\rm cm^2/s}$ at the stellar surface, which is close to the mass 
shedding limit. The rotation velocities of our model are slightly lower than those of 16TI in general. The 
specific angular momentum distribution on the equatorial plane as a function of enclosed mass is shown in 
Figure~\ref{f2}. Here the enclosed mass is defined as a mass within a certain radius. 
Presented also in this figure is the specific angular momenta at the innermost stable circular orbit (ISCO) 
for Schwartzshicld black holes as a function of their masses. The two curves intersects with each other at 
the enclosed mass of $\sim 8M_{\sun}$. Envelope matter that has a larger enclosed mass than this value can not 
fall down to the black hole and is halted somewhere outside the ISCO by centrifugal forces. As a matter of fact, 
we find a centrifugal bounce and a formation and subsequent propagation of a shock wave (see 
subsection~\ref{sec2.4}).

\subsection{Special Relativistic Hydrodynamic Code}
 Before proceeding to the subsequent steps, we describe the numerical methods employed for the hydrodynamic 
simulations done in Steps 2 to 4. We employ a 2D axisymmetric, special relativistic hydrodynamics code. 
Equatorial symmetry is also assumed in this paper. The basic equations we solve in this study are given 
as follows in the geometrical units $G = c = 1$, where $G$ and $c$ are the gravitational constant and speed of light:
\begin{eqnarray}
&&\partial_{t} \rho_{*} + \partial_{j}\left({\rho_{*} v^j}\right) = 0,  \label{eq:hydi}\\
&&\partial_{t} S_{r} + \partial_{j}\left( r^2 \sin{\theta} \hspace{1mm} T^{j}_{\hspace{1mm}r}\right)
 = r^2 \sin{\theta} \biggl\{ - T^{00} \hspace{1mm} \psi_{,r} + r \hspace{1mm} T^{\theta \theta}
 + r \sin^2{\theta} \hspace{1mm} T^{\phi \phi} \biggr\},  \\
&&\partial_{t} S_{\theta} + \partial_{j}\left( r^2 \sin{\theta} \hspace{1mm} T^{j}_{\hspace{1mm}\theta}\right) = r^2 \sin{\theta} \biggl\{ - T^{00} \psi_{,\theta}
 + r^2 \sin{\theta} \cos{\theta} \hspace{1mm} T^{\phi \phi} \biggr\},\\
&&\partial_{t} S_{\phi} + \partial_{j}\left( r^2 \sin{\theta} \hspace{1mm} T^{j}_{\hspace{1mm}\phi}\right)
 = 0, \\
&&\partial_{t} \tau + \partial_{j}\left( r^2 \sin{\theta} \hspace{1mm} T^{0j} - \rho_{*} v^j \right)
 =  - r^2 \sin{\theta} \hspace{1mm} T^{0i} \psi_{,i},   \\
&&\partial_{t} \left( \rho_{*} A \right) + \partial_{j}\left({\rho_{*} A v^j}\right) = 0,  \label{eq:hydf}\\
&& \Delta \psi = 4 \pi \rho_{0}
 \biggl\{ 2 h \left( u^t  \right)^2 - h + 2 \frac{p}{\rho_0} \biggr\},\label{eq:pot}
\end{eqnarray}
where  the subscript $j$ runs over $r$ and $\theta$, and $A$, $T^{\mu \nu}$, $u^{\mu}$ and $\psi$ 
denote the mean molecular weight, energy-momentum tensor of ideal fluid, four-velocity of matter and gravitational potential, respectively, and 
\begin{eqnarray}
\rho_{*} \equiv r^2 \sin{\theta} \rho_{0} u^{t}, \\
S_{i} \equiv r^2 \sin{\theta} T^{0}_{\hspace{1mm} i}, \\
\tau \equiv r^2 \sin{\theta} T^{00} - \rho_{*}.
\end{eqnarray}
The above equations are derived from the Einstein equations and energy-momentum conservation equations by 
the weak field approximation, ignoring the time-derivative of gravitational potential and space-derivatives
of three dimensional space metric. Since our computational domains do not contain the origin, the gravity of the 
central object is added as a point mass at the center. The time evolution of the mass of the central object is 
taken into account by integrating the mass flux crossing the inner boundary of the computational domain. 

We solve the Poisson equation for the gravitational potential, Eq.~(\ref{eq:pot}), by MICCG and the hydrodynamical 
equations, Eqs.~(\ref{eq:hydi})-(\ref{eq:hydf}), by the central scheme \citep{Kurganov2000,2008ApJ...689..391N}. 
In the latter, using the PPM interpolation method and TVD Runge-Kutta time integration, we achieve the second 
order accuracy in both space and time.  

The EOS's employed in this paper are the following. For Steps~2 and~3, that is, the construction of the 
stellar wind and the computation of the envelope collapse, the EOS by \citet{1996ApJS..106..171B} is used, 
in which the temperature and mean molecular weight are introduced to avoid the inconsistency with Step~1, 
where they are also accounted for. On the other hand, the so-called $\gamma$-law EOS, $p = (\gamma - 1) \rho_{0} 
\epsilon$ with $p$, $\gamma = 4/3$, $\rho_{0}$ and $\epsilon$ being the pressure, adiabatic index, rest mass 
density and specific internal energy, respectively, is adopted for the jet simulations in Step 4 for simplicity. 
Since we find that the envelope is radiation-dominated at the time of jet launch, this is a good approximation.

It is a consensus that high resolution simulations are necessary for the investigation of interactions 
between the jet and stellar envelope in the jet drilling phase, since the Kelvin-Helmholtz instability and 
turbulent motions take place inevitably. When the velocity of jet head is smaller than the local sound speed 
at the hot spot, which is indeed the case for the jet propagations in the stellar envelope, a back flow is bent and 
pinches the jet path \citep{2010ApJ...709L..83M}. In order to treat these effects adequately, we employ an 
adaptive mesh refinement (AMR) technique, in which the forward shock is searched at each time step and the 
number of mesh points in its vicinity is increased in each coordinate direction.

In Appendix, we show the results of several numerical tests meant to validate our hydrodynamics code used 
in this paper. It is also demonstrated that the rotational massive stellar envelope, which is constructed by
HSCF scheme at Step1, does not change the configurations in a dynamical simulation, which is clear evidence
the both codes are reliable.

\subsection{A Quasi-Steady Wind}
Massive stars experience mass losses in general and the GRB progenitors will not be 
exceptions \citep{2006Natur.442.1008C,2008Natur.454..246S}. We hence take into account the stellar wind in our 
initial model (Step2). It is noted, however, that theoretical understanding of the driving mechanism of stellar 
winds and mass losses of massive stars is far from satisfactory and it is much beyond the scope of this paper 
to address these issues. We are, therefore, satisfied with the construction of quasi-steady winds, not specifying 
its driving mechanism. It is stressed that what is important here is that the wind thus obtained does not 
change its configuration very much before the jet reaches it.

We first construct a spherically symmetric, steady wind configuration, neglecting rotation. For this purpose,
we perform 1D hydrodynamical simulation, using the code described above, in the region from the stellar 
surface up to the distance of $10^{13}$cm. The initial configuration is rather arbitrary. Fixing the density, 
pressure, velocity (or mass loss rate) at the inner boundary, we run a long-term simulation until the wind is 
settled to a steady state. The values of the density and pressure are chosen so that they would be continuous
when the wind is appended to the envelope model constructed at Step~1. Rotation is then added so that the 
specific angular momentum should be constant along each radial ray. The values of the specific angular momentum 
at the inner boundary are chosen in such a way that they would be continuous from the envelope to the wind. 
The wind obtained in this way is not exactly steady any more. Although rotational, steady wind configurations 
could be obtained in the similar way, it turns out that the wind configuration does not change much during 
the jet propagation through the envelope and wind if rotation is added this way. We hence do not pursue further
elaboration in this paper. 

\begin{figure}
\vspace{15mm}
\epsscale{0.5}
\plotone{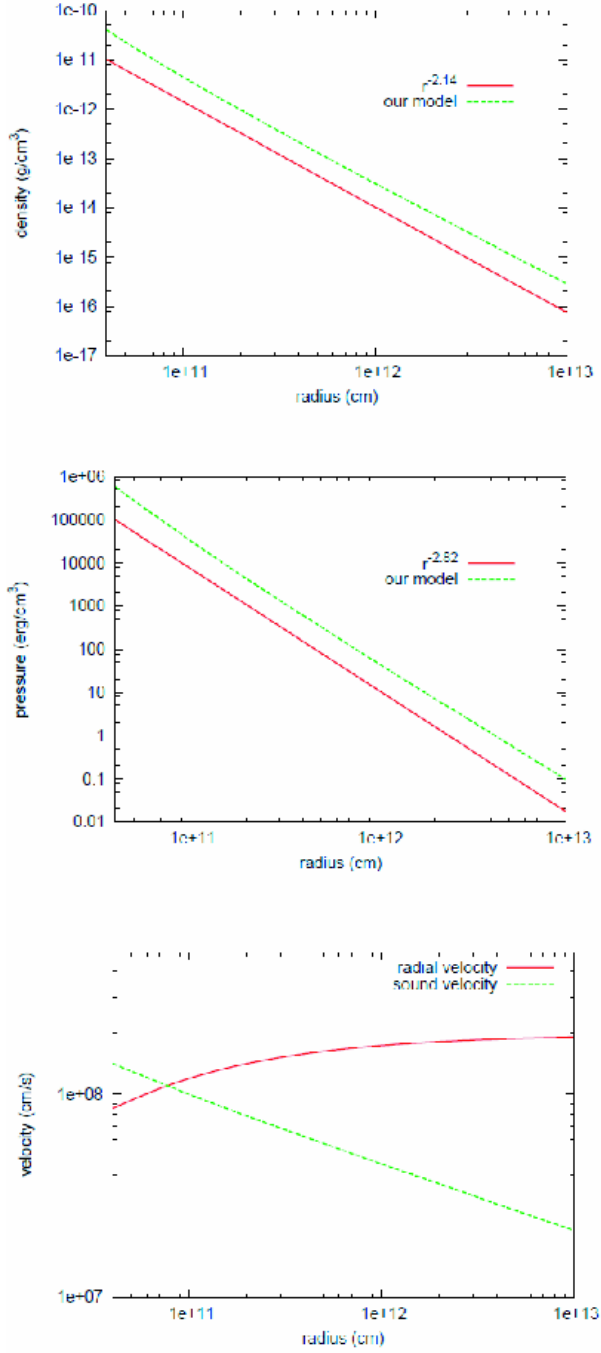}
\caption{ The profiles of density (upper), pressure (middle), and radial velocity as well as sound velocity (bottom) for the wind model
employed in this paper. The red lines in the upper and middle panels show power-laws for comparison. As shown in bottom panel, although 
the radial velocity is initially subsonic, it passes a sonic surface, which is located at $r \sim 7.5 \times 10^{10} {\rm cm}$.
\label{f3}}
\end{figure}

By changing the inner boundary condition, we can construct various wind models, both dense and tenuous. 
In this paper, however, we adopt only an optically thin model to elucidate the effects of 
envelope motions on the jet dynamics. Other wind models and their influences on the jet propagation will be 
investigated in the sequel to this paper \citep{Nagakura2011}. The photosphere of the present wind model is 
located at the stellar surface and the mass loss rate is $\dot{M} \sim 10^{-6} M_{\sun}$/yr. 
Figure~\ref{f3} shows the profiles of our wind model. The density and pressure distributions 
obey power laws approximately with the power-law indices being $-2.14$ and $-2.82$, respectively. The outflow 
in the wind becomes supersonic at $r \sim 7.5 \times 10^{10} {\rm cm}$ and its velocity approaches 
asymptotically $ v^{r}_{asym} \sim 2 \times 10^{8} {\rm cm/s}$.

\subsection{Collapse of the Massive Star Envelope \label{sec2.4}}
Using the envelope and wind configurations obtained above as an initial condition, we perform 2D axisymmetric 
simulations of the envelope collapse (Step3). The computational domain covers at first a region of 
$ 5 \times 10^6 {\rm cm} < r < 2 \times 10^{12}$cm, which includes the entire envelope and the core region except the 
black hole and its close vicinity as well as the inner part of the wind. The inner boundary is shifted outwards 
later (see below). The radial grid consisting of 224 points is non-uniform, with the grid width changing with the 
density scale height. The angular grid covers a quadrant of the meridian section and is uniform with 60 points. 

In reality, as we mentioned earlier, the gravitational collapse of the envelope is initiated by the arrival 
of the rarefaction wave that is generated at the core/envelope boundary by core collapse and propagates outward. 
To mimic this, we reduce the radial gradients of all quantities to zero at the inner boundary and artificially 
induce the infall there. Then a rarefaction wave is produced at the inner boundary and propagates outward, 
inducing infall at points it reaches. It is stressed that we confirmed by the test computation presented in 
Appendix \ref{apeq} that if we do not reduce the radial gradients of quantities at the inner boundary, the 
envelope remains intact even after many time steps.

As shown in the next section, the contraction of the envelope is eventually terminated by centrifugal forces, 
producing a shock wave that propagates outwards and eventually breaks out of the stellar envelope. We increase the 
number of radial grid points to $1000$ at the time of the shock breakout and shift the inner boundary outwards to 
$5 \times 10^8$cm simultaneously. All the quantities are linearly interpolated to the new mesh points. The change of
the inner boundary leads to the increase of the mass of the central object, which is properly taken into account,
whereas we discard the angular momentum and energy between the old and new inner boundaries just for simplicity.

It should be noted that our numerical code does not take into account general relativity and detailed microphysics 
such as photo-dissociations of nuclei and neutrino cooling. The neglect of these effects tends to overestimate 
the strength of the shock wave of centrifugal bounce-origin. In fact, it was pointed out by 
\citep{2010ApJ...713..800L,2010arXiv1007.0763M} that the nuclear photo-dissociations may completely sap the shock wave. We will
defer the investigation of this issue to a future work, in which we will implement a nuclear network in our 
hydordynamics code. It is also repeated that we assume in this paper that the prompt shock wave of core 
bounce-origin is swallowed into the central black hole and what occurs inside it does not affect the dynamics of 
the envelope. In order to see if this assumption is correct or not, it is necessary to perform detailed 
simulations of core collapse in full general relativity, which is a major undertaking and will also be a future 
work.

\subsection{Jet Injection and Propagations through the Stellar Envelope and Wind}

 In the next step (Step4), which is the main part in this paper, we numerically study the jet propagations 
through the stellar envelope and wind that are in motion as obtained in the previous step. Following the 
common practice, we inject a relativistic jet from the inner boundary, not specifying the driving mechanism, 
at two different times after the envelope collapse takes place: 20s for model M20s  and 50s for model M50s. 
The injection parameters are identical for both models: the jet is hot ($p/\rho_{0} c^2 = 20$, where $c$ is the 
speed of light) and relativistic with a Lorentz factor of 5; the half opening angle is $9^\circ$; the power 
of jet is constant in time and the injection continues for $t_{dur} = 30$s with the total injected energy being 
$10^{53}$ergs. Then the terminal Lorentz factor is estimated by
\begin{eqnarray}
\Gamma_{term} \equiv h_{in} \Gamma_{in} = (1 + \epsilon_{in} + p_{in}/(\rho_{in} c^2)) \Gamma_{in}
 \sim \gamma/(\gamma -1) \times p_{in}/(\rho_{in} c^2) \times \Gamma_{in}, \nonumber \\
 = 4 \{ p_{in}/(\rho_{in} c^2) \} \Gamma_{in},
\end{eqnarray}
where $h_{in}$, $p_{in}$, $\rho_{in}$ and $\Gamma_{in}$ are the specific enthalpy, pressure, rest mass 
density and Lorentz factor at the injection; the adiabatic index is denoted by $\gamma$ and is set to be $4/3$. 
The choice of the injection parameters in this paper corresponds to $\Gamma_{term} \sim 400$.

The computational domain for these simulations ranges from $r = 10^{9}$cm up tp $r = 10^{18}$cm. Note that this broad 
range is mandatory for the identification of the location of photosphere until $t_{obs} \sim 100$s, since the 
forward shock in the jet is highly relativistic with a Lorenz factor of $\Gamma > 100$. The total number of 
radial grid points is 11000. The grid is nonuniform, with the grid width being smallest ($\Delta r = 10^8$cm) at 
the inner boundary and increasing geometrically by $\sim 0.1 \%$ per zone up to $10^{13}$cm and by $\sim 1.35 \%$ 
in the region further out. The number of angular grid points is the same, 60, as in the previous step.
We remap the data obtained in the previous step to the new grid by the same linear interpolation as employed 
in Step3. The shift of the inner boundary requires an adjustment of the mass of central objects, with the 
mass between
 $ 5 \times 10^{6} \hspace{1mm} (5 \times 10^{8}) {\rm cm} < r < 10^{9} {\rm cm}$ 
being added to the central point mass for model M20s (M50s). The density, pressure and velocity in the region 
of $ 10^{13}$cm to $ 10^{18}$cm are extrapolated from the inner region in the following manner: The density and 
pressure are extended by the power-laws that fit their distributions in the inner region; The radial velocity is 
assumed to be constant in the extended region, since it has already reached the asymptotic velocity (see the 
bottom panel of Figure~\ref{f3}); The $\theta$ component of velocity is set to be 0, whereas 
the azimuthal component is determined so that the specific angular momentum is constant along each radial ray just
as in Step 2.

During the jet propagation through the stellar envelope, we employ an AMR technique. In our code there are only 
two levels of meshes deployed, in which the resolution of the second level can be varied. Here the mesh of the 
second level is 9 times finer than the first level mesh, with the smallest 
radial and angular resolutions being $ \Delta r = 1.1 \times 10^7$cm and $\Delta \theta = 0.16^\circ$, respectively. 
After the jet breakout, on the other hand, the jet head expands nearly freely and soon becomes highly 
relativistic. As a result, the back flow tends to be suppressed and the jet morphology does not change so much 
during this phase. We hence employ only 3 times finer a mesh for the second level after the jet head reaches $R = 10^{11} {\rm cm}$.
The resolution in this study is not as high as in the previous study \citep{2009ApJ...700L..47L}. One of the
main reasons for this is the fact that we are dealing with a much greater spatial extent. This is necessary, as 
already mentioned, to identify the locations of photosphere. As a result, however, rapid variations in the 
photospheric emissions are sacrificed to some extent by numerical dissipations and our discussions on this issue 
are restricted to a qualitative level. 



\subsection{Photospheric Emissions}
As a final step (Step5), we calculate, as a post process, the photospheric emissions based on the data 
obtained in Step4. We define the photosphere to be the surface that has a  unit optical depth from infinity with 
respect to the Thomson scattering. The optical depth is given by  
\begin{eqnarray}
 \tau (t_{obs}, r)  = \int ^{\infty}_{r} n_e(t^*, s) \, \sigma_{\rm T}  \, \Gamma (t^*, s)
  (1 - \beta(t^*, s) \cos \theta _v(t^*, s)) ds ,
\label{depth}
\end{eqnarray}
 where $s$ is the distance along the line of sight, 
 $n_e$ is the number density of electron in the comoving frame, $\sigma_{\rm T}$ is the cross section of the 
Thompson scattering, $\beta$ is the matter velocity normalized by the speed of light $c$,
$\Gamma$ is the corresponding bulk Lorentz factor and $\theta_v$ is the angle between the line of sight and the 
matter velocity \citep{1991ApJ...369..175A}.
 It should be stressed that the time retardation expressed by $t^* = t_{obs}-s/c$ in the above equation cannot be 
ignored for relativistic flows. In this paper, we evaluate Eq.~(\ref{depth}) as it is, retrieving the data 
for appropriate times from the results of the hydrodynamical simulations. Thanks to the wide spatial range of our 
hydrodynamic simulations, photons observed at $t_{obs} \lesssim 100$s have passed the forward 
shock by the end of simulations, the fact which is important for the identification of the locations of
photosphere.

 The observed isotropic luminosity of photospheric emissions is then given by
\begin{eqnarray}
 L = 
  4 \pi \int {\cal D}^4 
  I  \cos \theta_{\rm ph} \, dS .
\label{luminosity}
\end{eqnarray}
 Here $dS$ is the areal element of the photosphere (measured in the laboratory frame), 
 ${\cal D} = [\Gamma (1 - \beta \cos \theta_v)]^{-1}$ is the Doppler factor, 
 $\theta_{\rm ph}$ is the angle between the line of sight and
 the normal vector of the photosphere, 
 $I = \sigma_{\rm SB} \, T^4 / \pi$ is the radiation intensity with 
 $\sigma_{\rm SB}$ and $T$ being the Stefan-Boltzmann's constant and
 the temperature in the comoving frame, respectively. We ignore the cosmological redshift in this study.

\section{Results}

\subsection{Envelope Collapse}\label{subsectionenvelopecollapse}

\begin{figure}
\vspace{15mm}
\epsscale{1.0}
\plotone{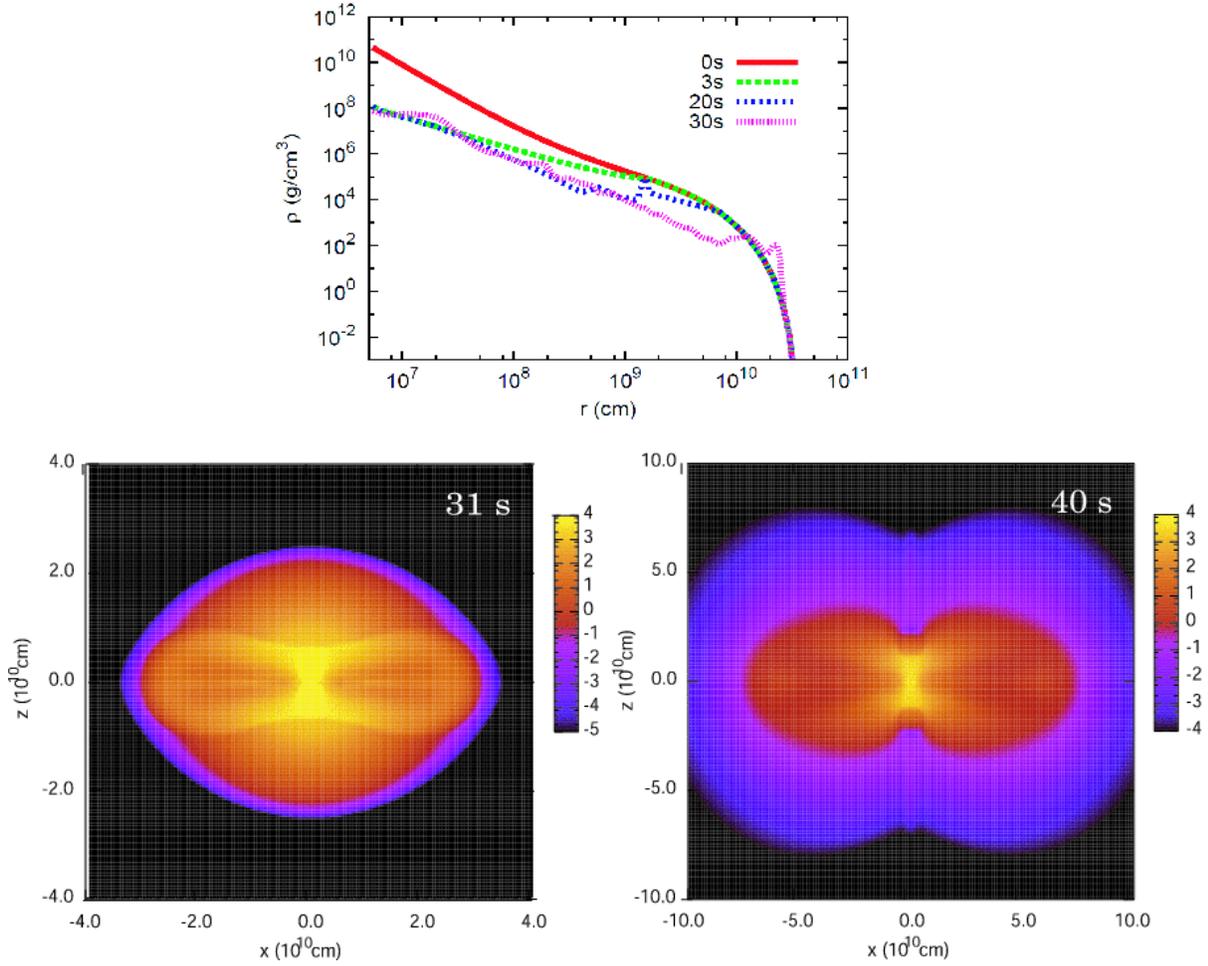}
\caption{ The time evolution of the density profile on the equator during the envelope collapse (top panel) and the density contours in 
the meridian section at the time of the breakout of the shock wave produced by centrifugal bounce (bottom left panel) and $\sim 10$s 
later (bottom right panel).
\label{f4}}
\end{figure}

The upper panel of Figure~\ref{f4} shows the temporal evolution of the density profile on the equator obtained in 
Step3, that is, the computation of envelope collapse. The infall starts at the inner boundary, generating a 
rarefaction wave that propagates outwards. Only after this rarefaction wave arrives, other parts of the envelope 
begin to move inwards. The contraction is almost spherical initially. As time passes and more distant portions of 
the envelope start to infall, however, the centrifugal force becomes non-negligible, since the specific angular 
momentum is an increasing function of radius. The centrifugal force becomes large enough eventually to halt the 
infall of matter and a shock wave is generated. This happens at $t \sim 18$~s in our model. A similar but a bit 
later bounce by centrifugal force was also reported by \citet{2010ApJ...713..800L}. The reason why we found the 
earlier formation of the shock wave is that we put the inner boundary at much smaller a radius than \citet{2010ApJ...713..800L}. Indeed, the inner boundary of our model is initially located at $\sim 3$ 
times the Schwarzschild radius, which is 10 times smaller than that adopted in \citet{2010ApJ...713..800L}. It 
should be noted that, as we have already mentioned, the shock wave is expected to be produced near the innermost 
stable circular orbit in reality and more accurate computation of its formation requires an implementation of 
general relativity as well as micophysics such as neutrino transports and photo-disintegrations of nuclei, which 
may sap the shock wave.

The shock wave propagates more vigorously along the equator than along the rotational axis and reaches the stellar surface on the equator at $t \sim 31$ s (see the bottom left panel of Figure~\ref{f4}). Then the shock wave breaks out of the stellar surface and runs further through the wind (see the bottom right panel of Figure~\ref{f4}). If the jet is launched earlier than the shock formation, the shock dynamics just described will be modified by the jet propagation. If the opposite is true, that is, the jet launch is later than the shock formation, the jet dynamics will be affected by the shock propagation. In particular, if the jet launch is sufficiently delayed, the jet propagation in the wind and, as a result, the photospheric emissions will be severely changed. Model M20s is meant for the former case whereas the latter case corresponds to Model M50s. Incidentally, the shock breakout in the latter case may account for the so-called precursor that is observed for some long
  GRBs. In fact, the typical time lag between the precursor and the prompt emission is several tens of seconds \citep{2005MNRAS.357..722L}, which is similar to what we find in our model. In this scenario, the high energy emissions in the precursor are supposed to be similar to those in the shock breakout of ordinary supernovae \citep{1978ApJ...225L.133F,1978ApJ...223L.109K,1999ApJ...510..379M,2007ApJ...667..351W,2008Natur.454..246S}. For more quantitative arguments for the precursor emissions from these shock waves are currently undertaken \citep{Nagakura2011}.

Incidentally, we assume in this paper that it is not the centrifugal bounce-originated shock wave but something 
else that is responsible for the jet launch. We hence treat the centrifugal bounce and the jet launch as 
independent events and vary the time of jet injection with respective to the centrifugal bounce rather freely. 
In reality, they may be correlated with each other one way or another. As we have already mentioned, 
the focus in this paper is the consequences that the possible time lag between these two events may have. 
The origin of the lag is intimately related with the mechanism of the jet launch. Although it is very 
interesting in its own right, the issue is much beyond the scope of this paper.



\subsection{Jet propagations in the Stellar Envelope and Wind}
\begin{figure}
\vspace{15mm}
\epsscale{0.8}
\plotone{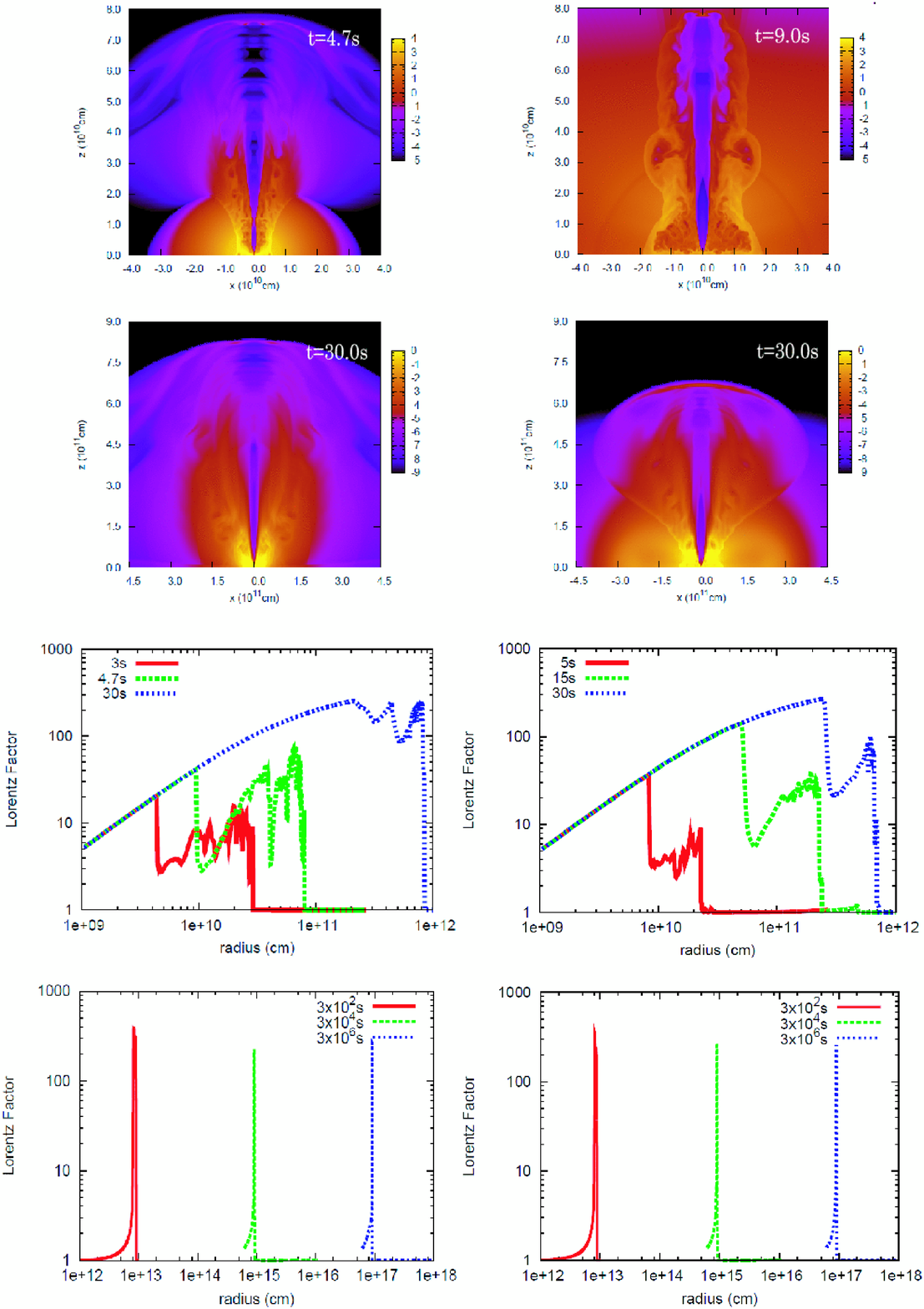}
\caption{ The density contours (upper four panels) and the time evolutions of the Lorentz factors on the rotational axis (lower four panels) 
for model M20s (left column) and model M50s (right column). The time, $t$, is measured from the instant, at which the jet is injected.
\label{f5}}
\end{figure}

As expected, the hydrodynamics of the early injection model, M20s, (see the left column of Figure~\ref{f5}) is 
similar to those found in the previous studies \citep{2009ApJ...700L..47L,2010arXiv1006.2440M}, since the envelope 
bounce by centrifugal force occurs almost at the same time as the jet launch and the outer envelope structure has 
not been changed very much from the initial one. The jet is strongly collimated by a hot cocoon, i.e. the shocked 
jet and envelope matter, until the jet breaks out of the progenitor surface. Then the shocked jet matter starts to 
expand laterally from the vicinity of rotation axis and the internal energy is gradually converted to the kinetic 
energy. As a result, the hot, shocked jet matter acquires a high Lorentz factor and produces very bright 
photospheric emissions (see next subsection). Since the jet injection is terminated at $t = 30$s in this model, a 
rarefaction wave is generated at that point and starts to chase the jet head and only the matter between the jet 
head and rarefaction wave contributes to subsequent radiations.



For model M50s, in which the jet is launched much later than the envelope bounce, the jet dynamics is very 
different from the one for the early injection case (see the right column of Figure~\ref{f5}). The jet propagates 
through the envelope that is not contracting but expanding owing to the shock wave produced at the centrifugal 
bounce of envelope. We find that the distance between the terminal (reverse) and forward shocks is shorter than 
for model M20s and the terminal shock remains to exist much longer for model M50s. The forward shock region in the 
jet is also found to be remarkably different after the breakout between the two models. Since the shock wave 
breaks out of the star before the relativistic jet reaches the stellar surface, the stellar wind is modified 
substantially by the shocked envelope matter (SEM). As a consequence, the jet propagation is hindered by the thick 
SEM even after its passing the position of the original stellar surface and the forward shock velocity becomes 
slower until much later times when the jet passes all through the SEM, producing a denser shell behind the 
forward shock (see the second panel in the right column of Figure~\ref{f5}). This has an important ramification 
for the photospheric emissions later on.

\subsection{Photospheric Emissions}

\begin{figure}
\vspace{15mm}
\epsscale{1.0}
\plotone{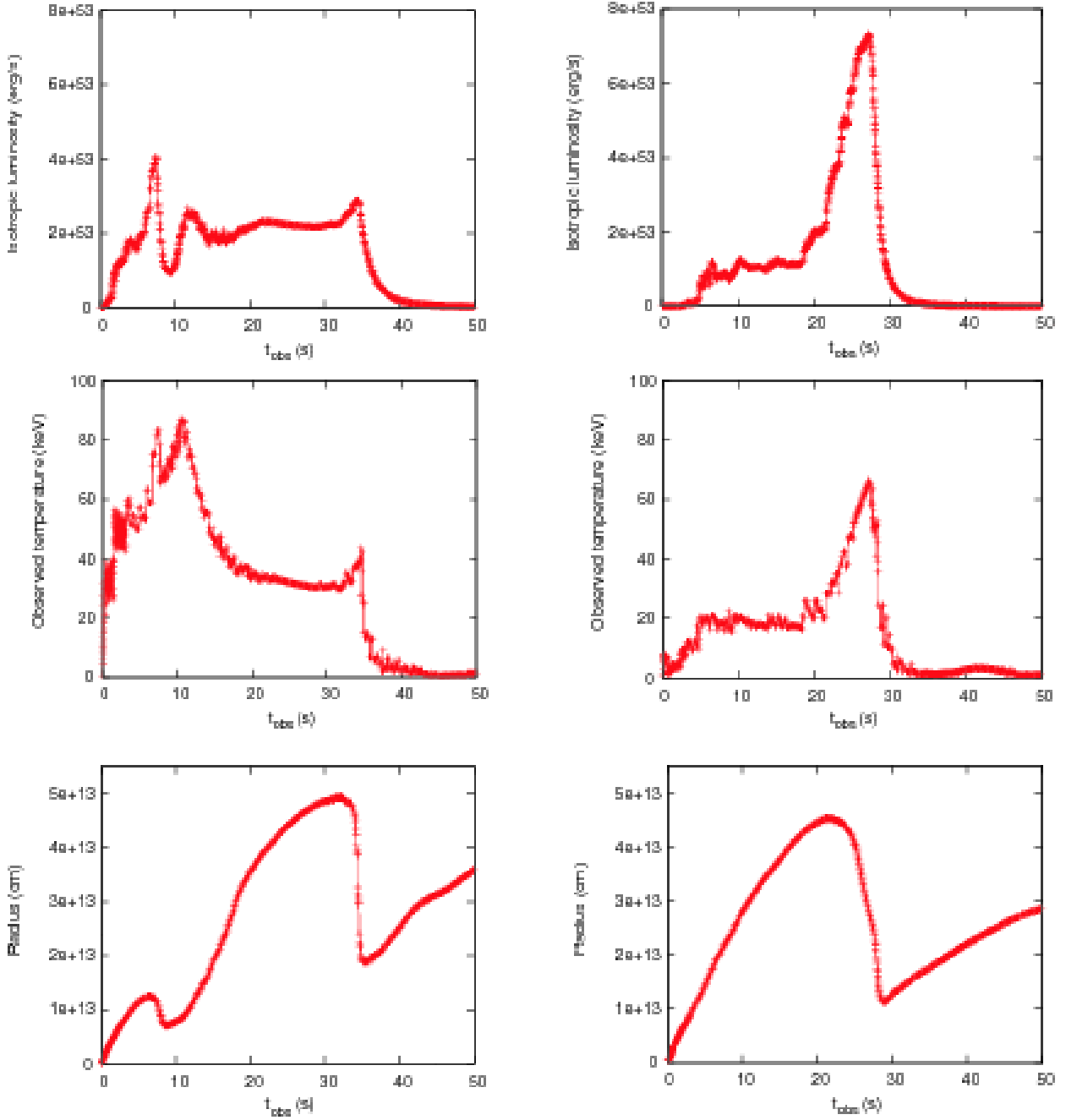}
\caption{ The light curves of photospheric emissions (top panels), the evolutions of the observed temperature (middle panels) and of the 
photospheric radius (bottom panels) as a function of the observed time for model M20s (left column) and model M50s (right column). See the 
body for details.
\label{f6}}
\end{figure}

In Figure~\ref{f6} we display the light curves together with the evolutions of photospheric radii and observed temperatures (${\cal D} T$) for both models. The observer is assumed to be located on the rotational axis (on-axis observer) and the photospheric radius in the figure is the value on the axis although we calculate the positions of photosphere for off-axis rays.
 For model M20s, the luminosity is peaked in an early phase ($t_{\rm obs} \lesssim 10{\rm s}$) and rather high luminosities are sustained for the next $30{\rm s}$ whereas a strong peak is observed at a late time ($t_{\rm obs} \sim 25{\rm s}$) for model M50s. As already mentioned, this difference arises from the large difference in the envelope structures prior to the jet breakout.
%
 Since a larger amount of matter is swept up by the jet in model M50s, it takes the photosphere longer to leave the forward shock region and move  inward to a region with higher observed temperatures, producing bright radiations. This qualitative difference in the light curves may be utilized to observationally extract information on the timing of jet launch at the central engine.

%

 As an explanation of the GRB prompt emissions, the photospheric emissions in our models have sufficiently high luminosities. The peak energy, however, is lower roughly by an order of magnitude than the value expected from the Yonetoku-relation \citep{2004ApJ...609..935Y}. This tendency is the main drawback of the photospheric emission model. Note, however, that the shocked jet matter may be scattering-dominant and energy exchanges between photons and matter may be terminated deeper inside. If this is the case, the observed temperatures will be higher and the luminosity will be also reduced. They are currently under study \citep{Ito2011}. It is also conceivable that some non-thermal processes are operating to produce high energy photons. Further exploration of these issues will require detailed computations of radiation transport and will also be a future work.

%

\section{Summary}
We have numerically investigated the propagations through a rapidly rotating massive star envelope of relativistic 
jets that are launched at different times, taking into account the motions of the envelope induced by core 
collapse. Then, we have calculated the photospheric emissions by post-processing. The main findings in this paper 
are summarized as follows:

1. In the envelope collapse, we have seen the generation of a shock wave by centrifugal force around $t \sim 20$s 
for the progenitor rotating uniformly at the mass shedding limit. In $\sim 10$s the shock wave breaks out of the 
star if the jet launch is sufficiently delayed.

2. If the shock wave produced by centrifugal force breaks out of the star earlier than the jet does, it changes the envelope and wind structures drastically and the jet propagation thereafter is also affected significantly. In fact, since the forward shock in the jet sweeps up a larger amount of matter, a dense shell is produced behind it in that case.

3. The light curve of photospheric emissions is qualitatively different if the jet is launched later and propagates in the shock-modified envelope and wind. In the case of earlier launch, the peak luminosity is attained at a relatively early time ($t_{obs}\sim 10$s), whereas it takes longer ($t_{obs}\sim 25$s) to observe the peak for the delayed launch case owing to the dense shell just mentioned.

4. The photospheric emissions obtained in this study with the time retardation being taken into account appropriately have high luminosities suitable for the GRB prompt emissions. However, the peak energy tends to be lower than expected from the Yonetoku-relation. If the shocked jet matter is scattering-dominated, photons will cease to exchange energy with matter deeper inside, where the temperature is higher. It is also possible that some non-thermal processes boost the photon energy.

The light curve of our results are different from those given in the previous papers 
\citet{2009ApJ...700L..47L,2010ApJ...709L..83M}. Since the focus in this paper is to investigate 
possible consequences that the difference in the timing of jet launch may have for the prompt emissions, 
we have not made detailed comparisons with these previous studies. There are a couple of conceivable 
causes for the differences: (1) better estimation of the location and temperature of photosphere thanks to
the wider computational domain, (2) the effect of envelope collapse taken into account, (3) the differences in
the jet-injection parameters, progenitor models and numerical resolutions. These issues will be addressed in 
our forthcoming paper.

\acknowledgments
 We thank Akira Mizuta and Shigehiro Nagataki for useful discussions. This work is partially supported by the Japan Society for Promotion of Science (JSPS) Research Fellowships, Grant-in-Aid for the Scientific Research from the Ministry of Education, Culture, Sports, Science and Technology, Japan [Nos. 222913, 22740178, 21540281, 19104006]. This study is also supported by Program for Improvement of Research Environment for Young Researchers from Special Coordination Funds for Promoting Science and Technology (SCF) commissioned by the Ministry of Education, Culture, Sports, Science and Technology (MEXT) of Japan.

\appendix

\section{Code Tests}
In this appendix, we carry out a series of tests in order to validate our special relativistic hydrodynamics code, which employs
the PPM reconstruction and TVD Runge-Kutta integration with a second order accuracy in both space and time. We adopt a HLL-type 
numerical flux and the CFL number is set to be 0.5. Included in the following are (1) 1D special relativistic shock tube 
problems, (2) the same as (1) but with tangential velocities and (3) a 2D Riemann problem. The 1D problems are compared with the exact 
solutions, and we utilize the results given in previous papers for the 2D problem. We also solve (4) one-dimensional and (5) 
two-dimensional isentropic flows to obtain the convergence rate quantitatively. For these test runs (1)-(5), the $\gamma$-law EOS is
adopted with the adiabatic index of $ \gamma = 5/3$. (6) We also run a dynamical simulation of the rotating stellar envelope, which is obtained 
by the HSCF method (see section~\ref{roteq}) and confirm that the stellar envelope sustains the initial profile for a long time. 
In order to check the accuracy of our AMR part, (7) we compute a non-relativistic, spherical point explosion, which can be compared 
with the Sedov-Taylor analytical solution, (8) a pulse propagating adiabatically through meshes of different refinement levels, and (9) 
an axisymmetric, relativistic jet propagation in a uniform matter. These tests demonstrate that our numerical code has enough accuracy for the purpose of the current study. Throughout this appendix, we adopt geometrical units $G = c = 1$ otherwise stated.

\subsection{1D Relativistic Shock Tube Problems Without Tangential Velocities}
The shock tube problem is one of the common tests for hydrodynamical codes. It is a special Riemann problem in gas dynamics. 
One of the advantages of this test is the fact that we know exact solutions even in special relativity \citep{2000JFM...422..125P}. 
We can check how well the code reproduces the profile of a rarefaction wave and captures several discontinuities such as contact surface 
and shock wave. In this test, we set the number of grid points to be 400 and the parameters employed for two runs are as follows:

case 1. Left state: $(\rho,v,p)^{L} = (10,0,13.3)$, \hspace {0.63cm} Right state: $(\rho,v,p)^{R} = (1,0,10^{-6})$

case 2. Left state: $(\rho,v,p)^{L} = (1,0,10^{3})$, \hspace {1cm} Right state: $(\rho,v,p)^{R} = (1,0,10^{-2})$

\begin{figure}
\vspace{15mm}
\epsscale{1.0}
\plotone{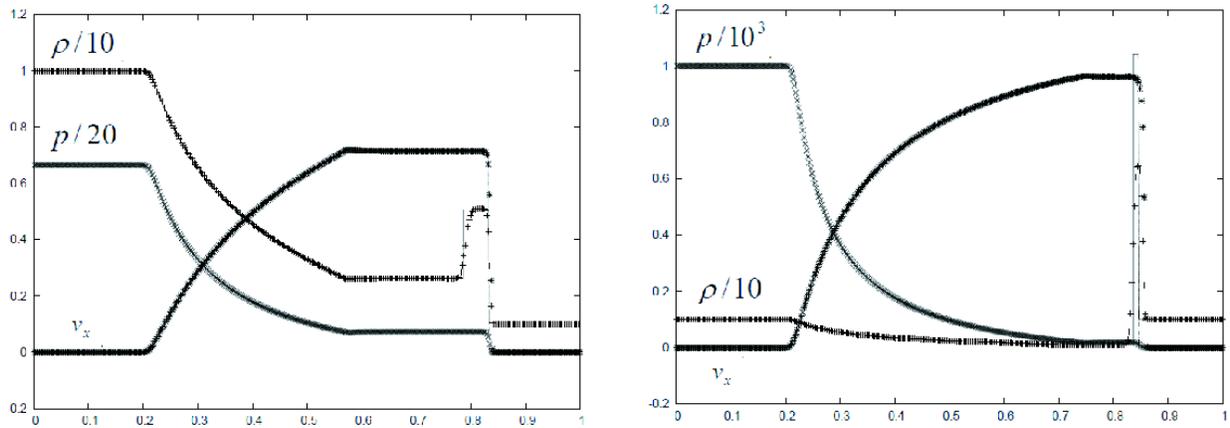}
\caption{Numerical results (dots) for the 1D relativistic shock tube problems without tangential velocities. The rest mass density ($\rho$),
pressure $p$ and velocity $v_x$ are shown. The exact solutions (solid lines) are also displayed for comparison.  The left (right) panel 
corresponds to case 1 (case 2) at $t = 0.4$ ($t = 0.35$).
\label{f7}}
\end{figure}

Figure~\ref{f7} shows the results at $t = 0.4$ together with the exact solutions. As is obvious, the overall profiles 
are well reproduced. Although the contact surface and shock wave are somewhat smeared out, our results are quite similar to those of 
other groups \citep[see e.g.][]{2002A&A...390.1177D}.

\subsection{1D Relativistic Shock Tube Problems With Tangential Velocities}
Here we show the results of relativistic shock tube problems with tangential velocities. In the special relativistic shock tube problems, 
the velocity components tangential to a discontinuity play a non-trivial role unlike in the non-relativistic counter part because the 
Lorentz factor depends on the absolute value of the velocity and it is numerically harder to resolve the flow profiles in special relativity 
as reported by \citep{2000JFM...422..125P,2002PhRvL..89k4501R}. In these tests, we adopt the same initial condition as in case 2 of the
previous subsection except for the non-vanishing tangential velocities, which are identical to those in \citet{2006ApJ...651..960M}.

\begin{figure}
\vspace{15mm}
\epsscale{1.0}
\plotone{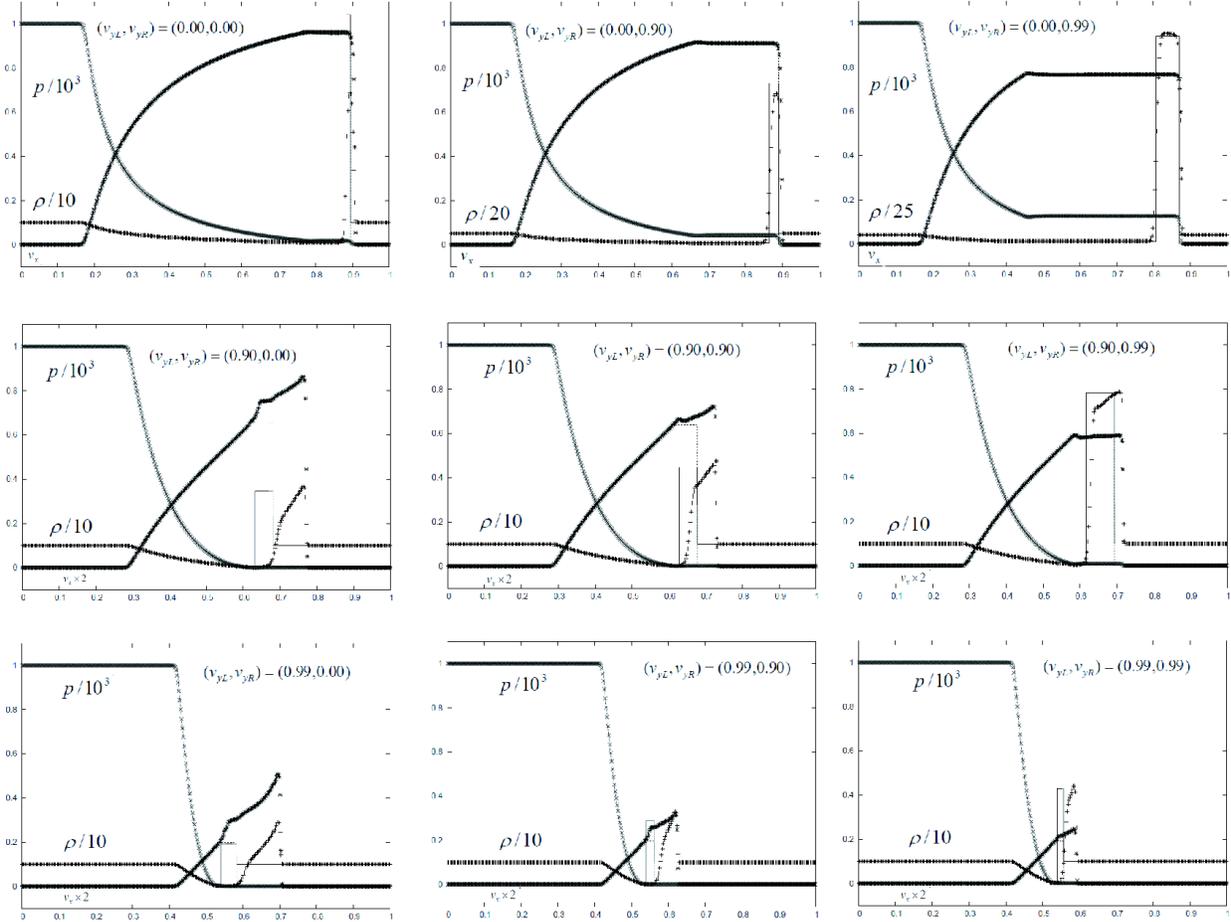}
\caption{Numerical results (dots) for the 1D relativistic shock tube problems with tangential velocities. A uniform mesh with 400 grid 
points are employed. The exact solutions (solid lines) are also displayed for comparison. We change $v_y^R$ from left to right as 
$v_y^R = 0, 0.9, 0.99$ and $v_y^L$ from top to bottom as $v_y^L = 0, 0.9, 0.99$. The density, pressure and x-component of velocity are 
shown. See Figure~15 in \cite{2006ApJ...651..960M} for comparison.
\label{f8}}
\end{figure}

\begin{figure}
\vspace{15mm}
\epsscale{1.0}
\plotone{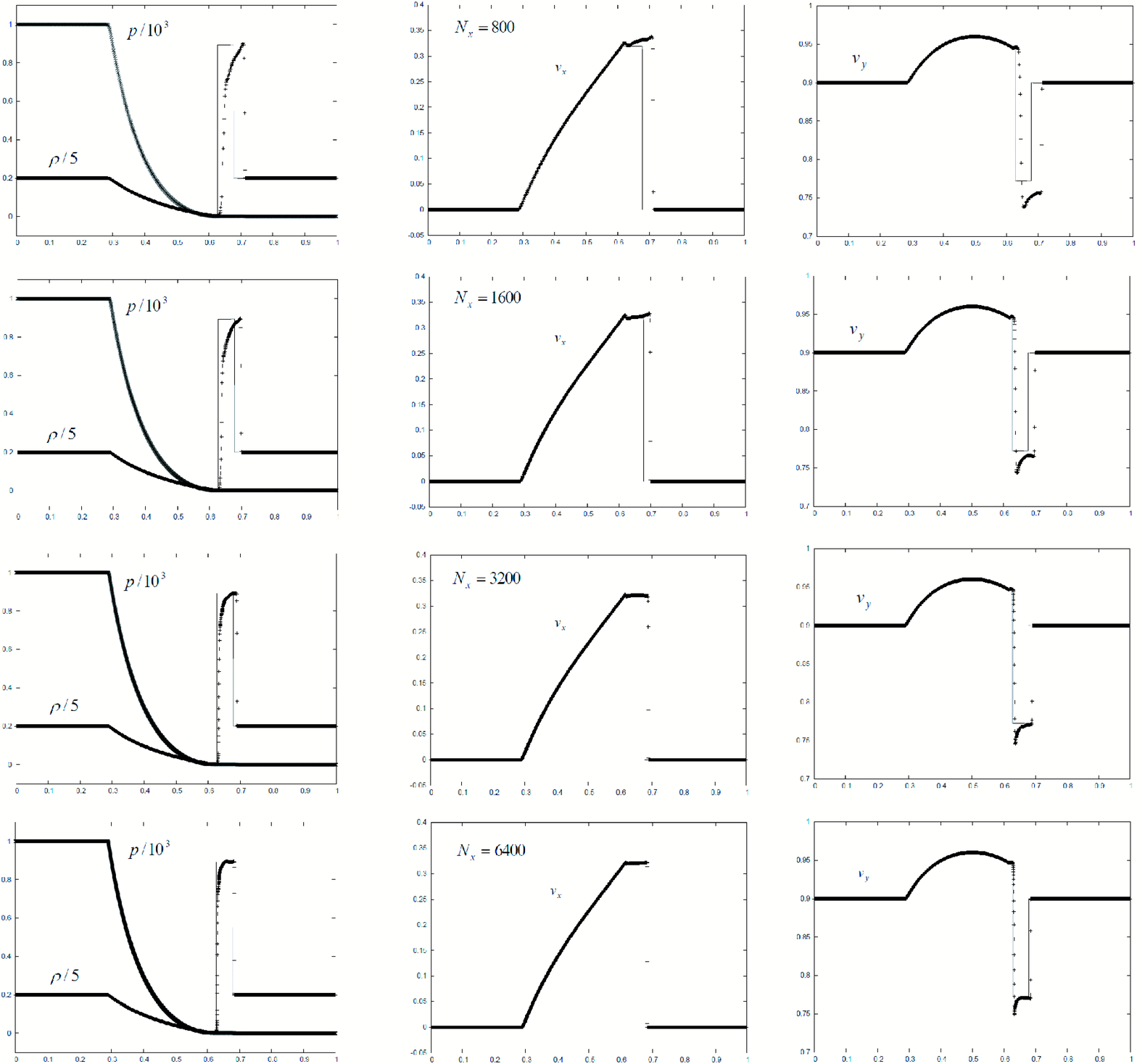}
\caption{Numerical results (dots) for the same problem as in the previous figure $(v_y^L,v_y^R)=(0.9,0.9)$ with different resolutions. These are meant to check the numerical convergence. The exact solutions (solid lines) are displayed for comparison. The left panels show the rest mass density and 
pressure, whereas the middle (right) panels display the x-component (y-component )of velocity. From top to bottom, the number of grid points 
are 800, 1600, 3200 and 6400, respectively.
\label{f9}}
\end{figure}

Figure~\ref{f8} shows the results of these tests. We have varied the tangential velocity $v_y$ from $0$ to $0.99$ on 
both sides. It is clear that both the contact surface and shock wave are substantially deviated from the exact solutions as the tangential 
velocity becomes large. We have performed test runs for $(v_y^L,v_y^R) = (0.9,0.9)$ with higher spatial resolutions (the number of 
grid points change from 800 to 6400) and display the results in Figure~\ref{f9}. Although the deviations of the 
numerical results from the exact solution are still noticeable even in these high resolution runs, we can confirm the convergence of 
the numerical results to the exact solution. Again the performance of our code is similar to others \citep[see e.g.][]{2006ApJ...651..960M,2006ApJS..164..255Z}.

\subsection{A 2D Riemann Problem}
This test is meant to check the performance of our code in multi-dimensional settings. The computational domain is initially divided 
in 4 sections, which have different states. The solution consists of multiple shock waves, contact surfaces and a rarefaction wave 
interacting with each other. The parameters we adopt in this test are the same as in \citet{2002A&A...390.1177D,2006ApJ...651..960M,2007ApJ...665..569M}:

$(\rho,v_x,v_y,p) = (0.1,0.00,,0.00,0.01)$ for $0.5 \ge x \ge 1$, $0.5 \ge y \ge 1$

$(\rho,v_x,v_y,p) = (0.1,0.99,,0.00,1.00)$ for $0 \ge x \ge 0.5$, $0.5 \ge y \ge 1$

$(\rho,v_x,v_y,p) = (0.5,0.00,,0.00,1.00)$ for $0 \ge x \ge 0.5$, $0 \ge y \ge 0.5$

$(\rho,v_x,v_y,p) = (0.1,0.00,,0.99,1.00)$ for $0.5 \ge x \ge 1$, $0 \ge y \ge 0.5$

\begin{figure}
\vspace{15mm}
\epsscale{1.0}
\plotone{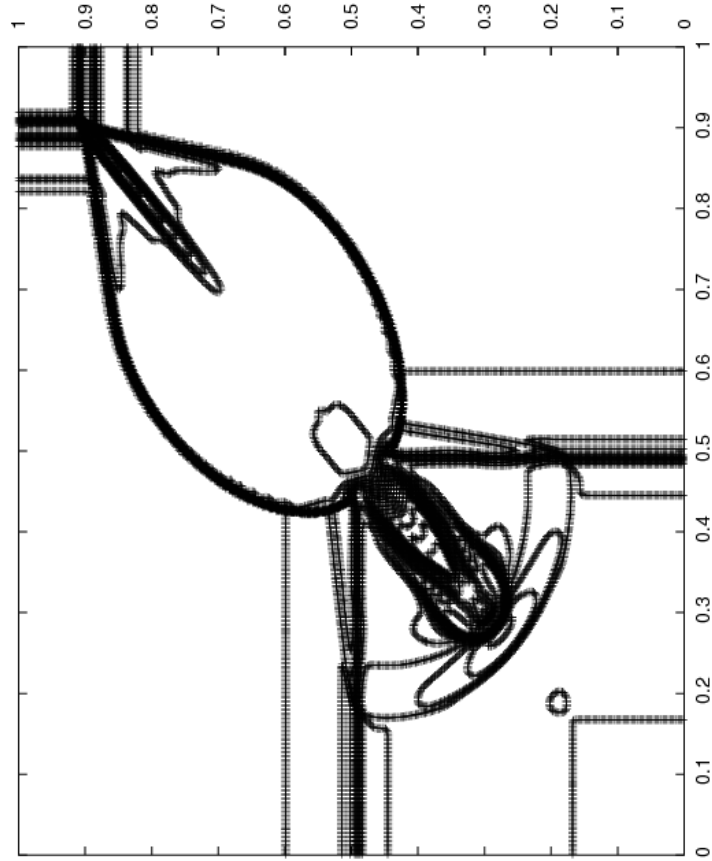}
\caption{The contour plot of rest mass density in the logarithmic scale at $t = 0.4$ for the 2D Riemann problem. 
\label{f10}}
\end{figure}

We use a uniform mesh with $400 \times 400$ grid points. Figure~\ref{f10} shows a contour in the logarithm scale of the 
rest mass density obtained in this simulation. There is no exact solution available. The results appear very similar to the ones presented 
in previous studies.

\subsection{A 1D Isentropic Flow} \label{subseconedimeisentflow}
All the above tests involve discontinuities such as shock wave and, as a result, the code affords only a first order accuracy because
the numerical error is dominated by these structures. Note that this is necessary to ensure numerical robustness. In order to see the 
performance of our code for smooth flows, we have carried out numerical simulations of 1D and 2D (see the next subsection) isentropic 
flows. The exact solutions are obtained by the characteristic method. The test hence offers an opportunity to quantitatively assess the 
accuracy of our code.

The initial condition for this test is the same as those employed in the previous studies \citep{2006ApJS..164..255Z,2007ApJ...665..569M} and the density 
profile is given by
\begin{eqnarray}
\rho_{0}(x) = \rho_{ref} \{ 1 + \alpha f(x) \}, \label{isentroinirhopro}
\end{eqnarray}
where $\rho_{ref}$ is the density of a reference state and
\begin{eqnarray}
f(x) = \left\{ \begin{array}
                  {l@{\quad:\quad}l}
                 ((x/L)^2 - 1)^4 & |x|<L  \\ 0 & \mathrm{otherwise}
                  \end{array} \right. \label{isentroinirhopro_f}
\end{eqnarray}
with $\alpha$ and $L$ being the amplitude and width of a pulse, respectively. Since the flow is isentropic, we employ a polytropic 
EOS ($p = K \rho_{0}^{\gamma}$) with the polytropic constant of $K = 100$ and the adiabatic index of $\gamma = 5/3$. The velocity of 
the reference state is set to be 0, while the velocity distribution inside the pulse is chosen so that the left-going Riemann invariant 
should be constant. Thanks to this set-up, the wave propagates in one direction. The special relativistic Riemann invariants are given 
by 
\begin{eqnarray}
  J_{\pm} = \frac{1}{2} \ln (\frac{1+v}{1-v}) \pm
  \frac{1}{\sqrt{\Gamma-1}}
  \ln (\frac{\sqrt{\Gamma-1}+c_s}{\sqrt{\Gamma-1}-c_s} ), \label{eqJpramin}
\end{eqnarray}
where $c_s$ denotes the sound velocity. The equations of characteristics $C_{\pm}$ are expressed as
\begin{eqnarray}
\left( \frac{dx}{dt}\right)_{C_{\pm}} = \frac{v \pm c_s}{1 \pm v c_s}.
\end{eqnarray}
Since the $J_{-}$ and entropy are constant over the whole region, the pulse evolution is determined by the $J_{+}$, which is carried along 
the characteristic $C_{+}$ until right-traveling characteristics collide with each other and a shock wave forms. Although the shape of 
pulse is initially symmetric, it is skewed owing to different characteristic velocities. Note also that the post-pulse state is the reference state.

\begin{figure}
\vspace{15mm}
\epsscale{0.5}
\plotone{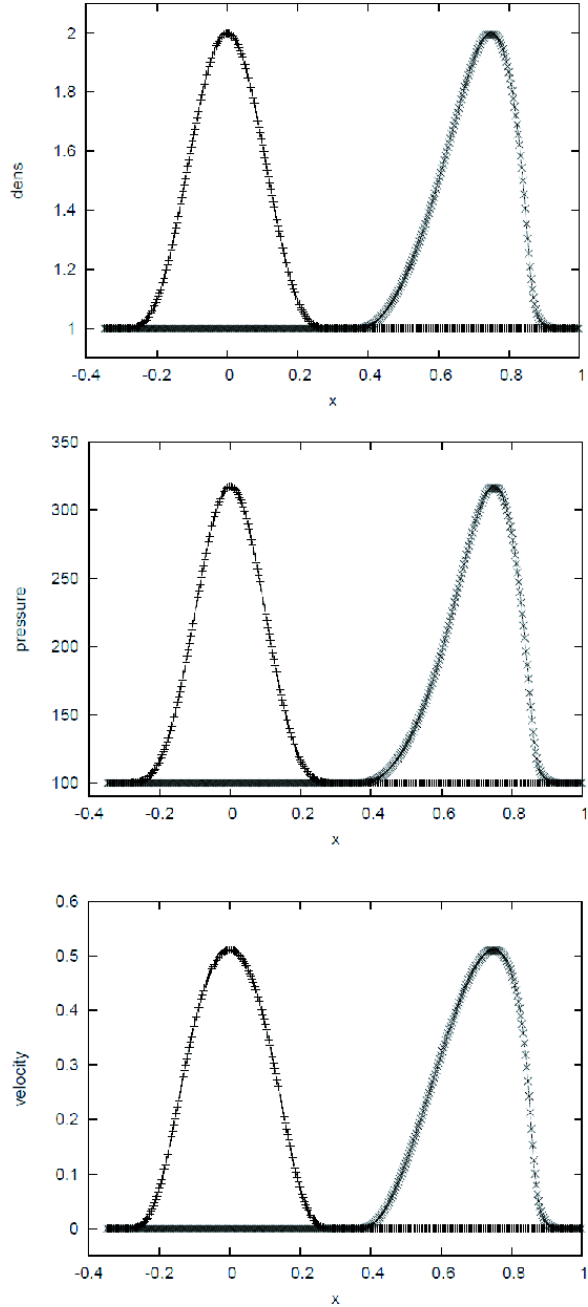}
\caption{The initial ($t = 0$) and simulated ($t = 0.8$) density (top), pressure (middle), velocity (bottom) profiles for the
1D isentropic flow together with the exact solution (solid lines). A uniform mesh with 400 grid points is employed. 
\label{f11}}
\end{figure}

\begin{figure}
\vspace{15mm}
\epsscale{1.0}
\plotone{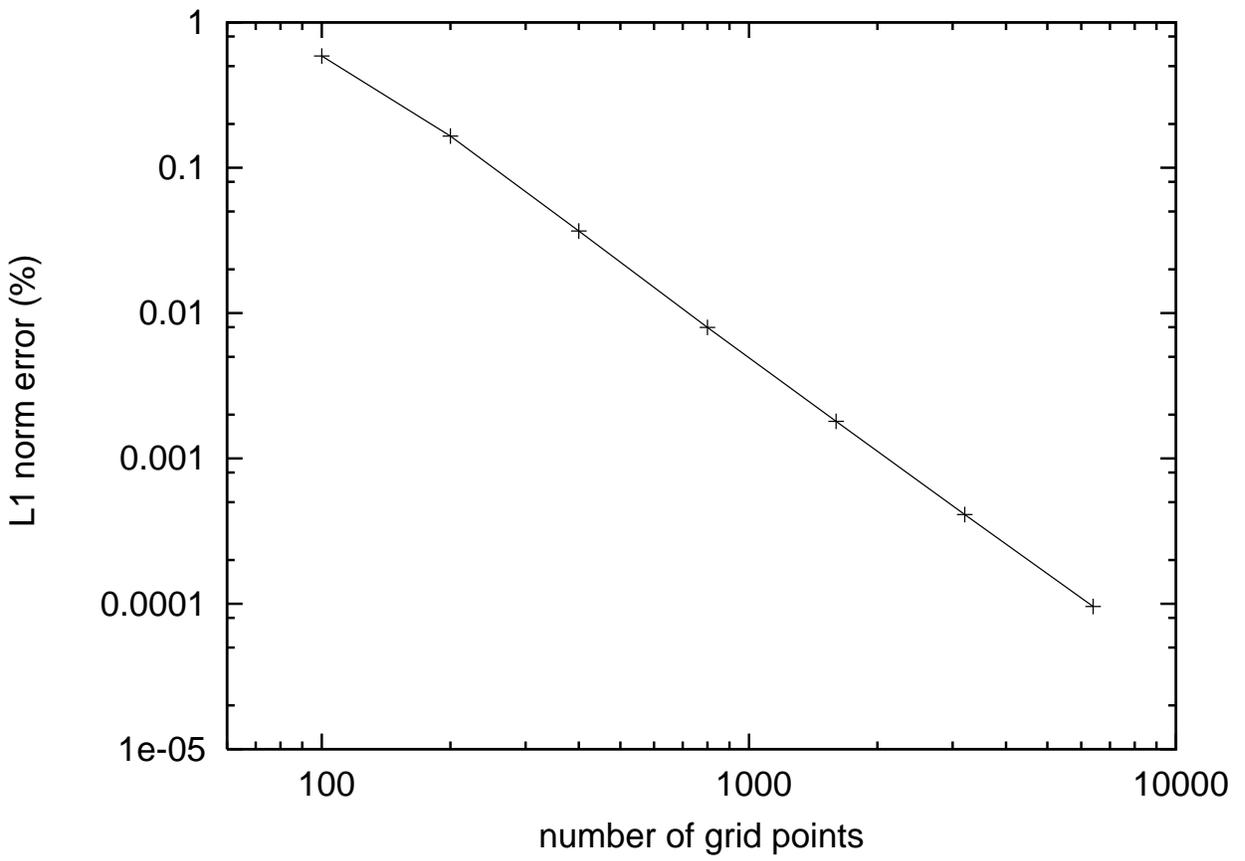}
\caption{The $L^1$ density errors for the 1D isentropic flow as a function of the number of grid points.
\label{f12}}
\end{figure}

Our computational domain is the same as in previous studies \citep{2006ApJS..164..255Z,2007ApJ...665..569M}: ($-0.35 \le x \le 1$); The 
reference state has $\rho_{ref} = 1$, $p_{ref} = 100$ and $v_{ref} = 0$; The amplitude of the pulse is $\alpha = 1.0$ and 
the width is $L = 0.3$. The simulation is run until $t = 0.8$. A comparison of numerical and exact solutions is displayed in Figure~\ref{f11}. We have also calculated the $L^1$-norm errors in density for 
different spatial resolutions, where $L^1 \equiv \Sigma_{j} \Delta x_{j} | \rho_{0j} - \rho_{0}(x_j)  | $ with $\rho_{0j}$ and 
$\rho_{0}(x_j)$ being the numerical and exact solutions, respectively. In Table~\ref{tab1}, we summarize the results of 
convergence check for this problem. It is thus confirmed that our code has indeed a second order convergence for smooth flows (see also Figure~\ref{f12}).

\subsection{A 2D Isentropic Flow}
We have also performed a 2D computation of the isentropic flow to assess the convergence rate of our code in a multi-dimensional context. 
The initial condition for this test is the same as in \citet{2006ApJS..164..255Z,2007ApJ...665..569M}. The computational region is a 2D 
Cartesian box with $0.0 \le x \le 3.75$ and $0.0 \le y \le 5.0$. The periodic boundary condition is adopted for all four sides of the box. 
The reference state is set to $\rho_{ref} = 1$, $p_{ref} = 100$ and $v_{ref} = 0$. The polytropic EOS with the polytropic constant of 
$K = 100$ and the adiabatic index of $\gamma = 5/3$ is again employed just as in the 1D case. Periodic pulses are prepared initially in such 
a way that they have a spatial period of $S = 3.0$ in the direction given by a unit vector, ${\bf k} = (4/5,3/5)$, and are uniform in the
perpendicular direction. The projected spatial periods on the x- and y-directions are 3.75 and 5.0, respectively, which is consistent 
with the size of our computational domain. The initial density profile is given by $\rho_{0} (d)$ 
(eqs. (\ref{isentroinirhopro}) and (\ref{isentroinirhopro_f})), in which $d$ is a distance from the center of the nearest pulse and expressed as 
$d = {\rm mod} ( {\bf k} \cdot {\bf r} + S/2,S ) - S/2$ with ${\rm mod}(a,b)$ being a function defined as 
${\rm mod}(a,b) \equiv a - \lfloor a/b \rfloor \times b$, where $ \lfloor a/b \rfloor$ denotes the integer part of $a/b$. The amplitude of 
the pulse is chosen to be $\alpha = 1.0$ and the width is set to be $L = 0.9$. The velocity distribution in the pulse is determined as in the
1D case in the previous subsection so that $J_{-}$ defined for this oblique 1D problem should be constant. The simulation is run up to $ t = 2.4$.

\begin{figure}
\vspace{15mm}
\epsscale{1.0}
\plotone{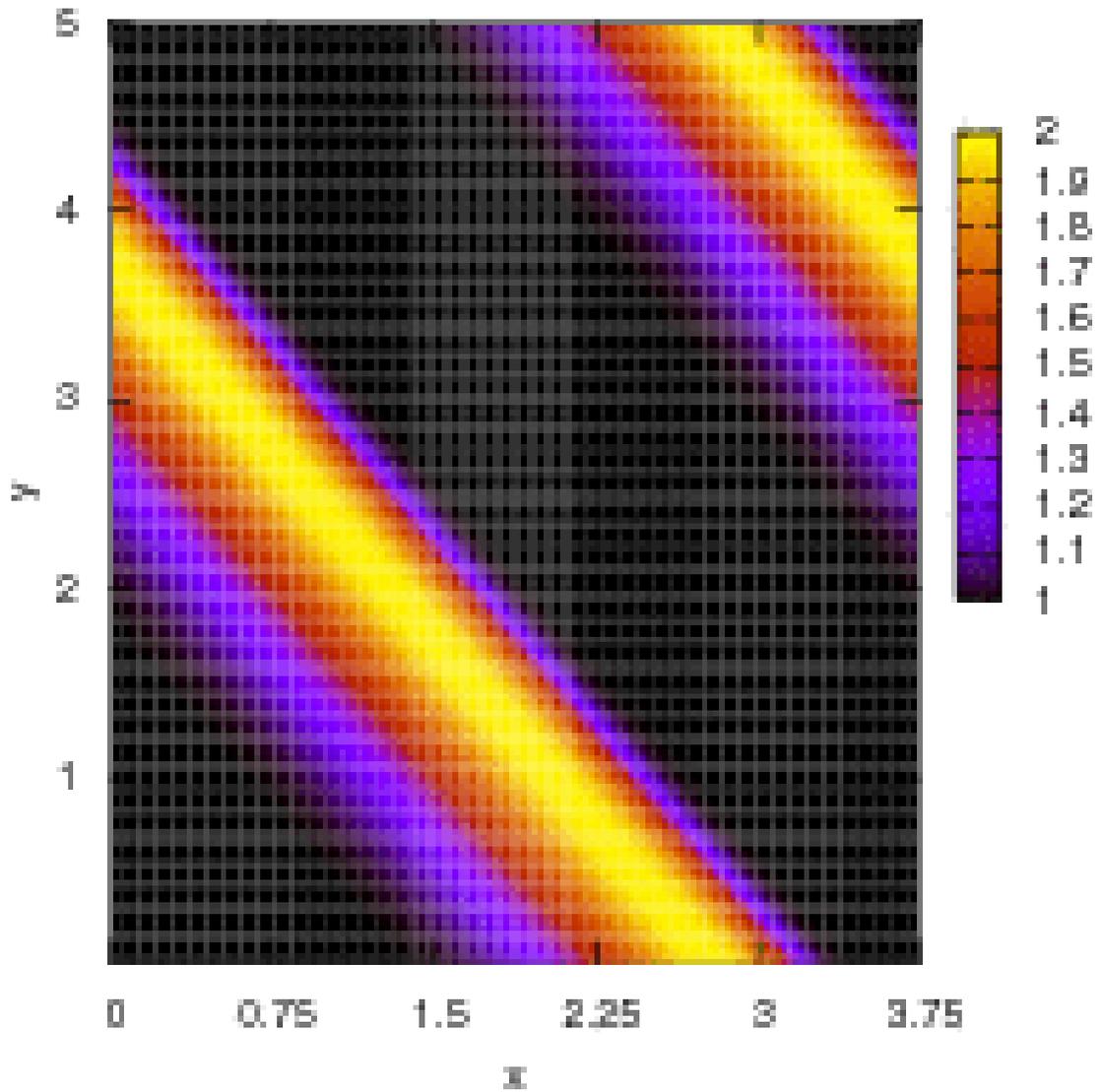}
\caption{The numerical result of the density structure for the 2D isentropic-flow problem. The spatial resolution is 
effectively equivalent to the numbers of grid points of (96,128) in the $x-$ and $y-$directions.
\label{f13}}
\end{figure}

\begin{figure}
\vspace{15mm}
\epsscale{1.0}
\plotone{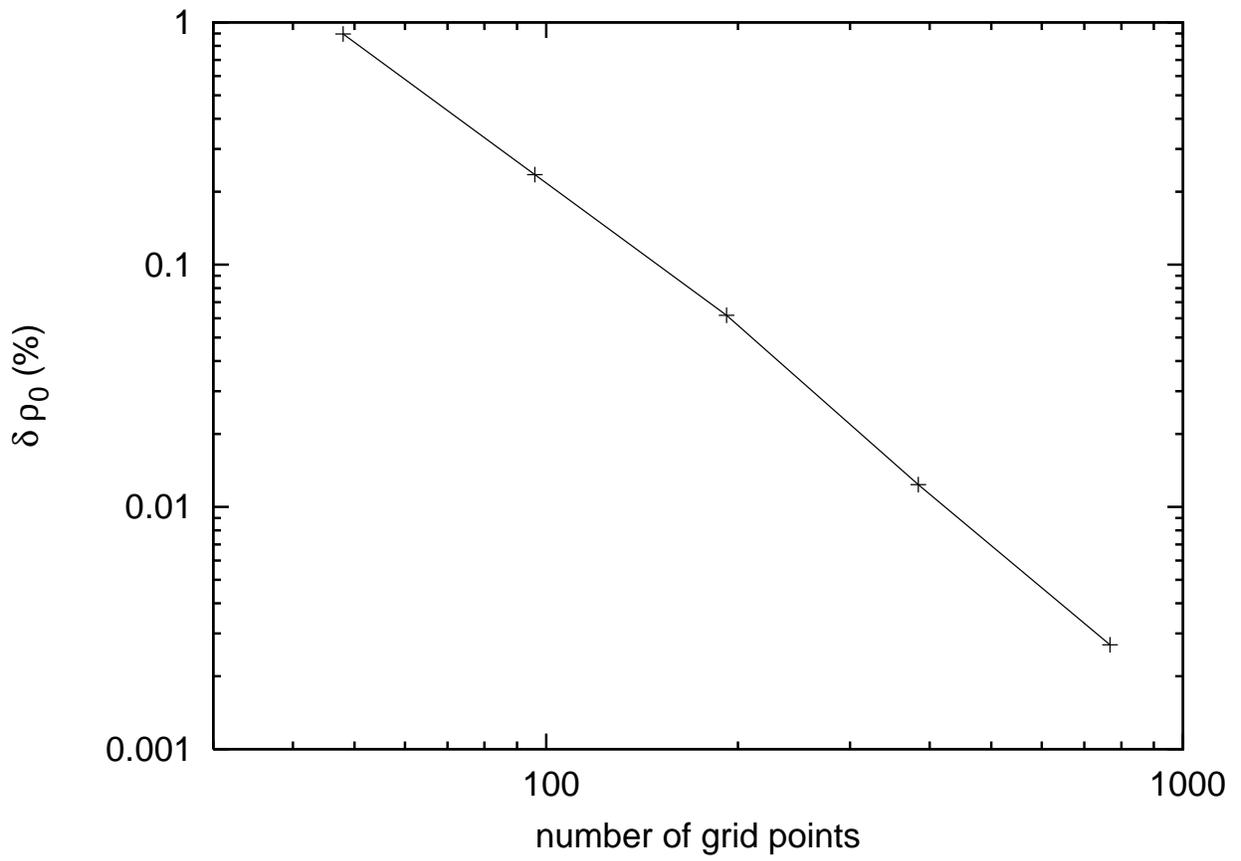}
\caption{The density errors $\delta \rho_{0}$ for the 2D isentropic flow as a function of the number of grid points.
\label{f14}}
\end{figure}

Figure~\ref{f13} shows the numerical result as a density contour at $t = 2.4$. In this figure, we display the case which the numbers of grid points are set to be 96 and 128 in $x-$ and $y-$direction, respectively. Just like the one dimensional counter part, the pulse becomes asymmetric with the right 
side of pulse becoming narrower than the left side. We have also run some simulations with different numerical resolutions and confirmed
again that the convergence is approximately of second order (see Table~\ref{tab2} and Figure~\ref{f14}). Note that the error, $\delta \rho_{0}$, is defined as an 
$L^1$ norm in 2D to be
\begin{eqnarray}
\delta \rho_{0} \equiv \frac{ \Delta x \Delta y \Sigma_{j,k} | \rho_{0jk} - \rho_{0}(x_{j,k})  |     }
{\Delta x \Delta y  \Sigma_{j,k}  \rho_{0}(x_{j,k}) }
\end{eqnarray}
where $\rho_{0jk}$ and $\rho_{0}(x_{j,k})$ are the numerical and exact solutions for density at the mesh point having an address $(j,k)$.

\subsection{Dynamical Simulations of Rotational Equilibrium}\label{apeq}
In order to check the consistency of the dynamical code with the code of the HSCF method employed to construct the rotational equilibrium,  
we have run long-term simulations for the stellar envelope in rotational equilibrium, which is obtained by the HSCF method (see 
section~\ref{roteq}). The initial configuration will not change in time if it is indeed in dynamical equilibrium. The test will hence 
validate both the hydrodynamics code with the weak gravitational field approximation and the HSCF code simultaneously. The computations 
are essentially the same as those done in Step 3 in the main body except for a different treatment of the boundary condition. All the 
quantities are fixed at the boundaries to the values provided by the HSCF calculations unlike in Step 3, where the radial gradients of them
are artificially reduced to zero. In this subsection, we adopt cgs units. The computational domain covers a radial extent of 
$10^8 {\rm cm} < r < 2 \times 10^{10} {\rm cm}$. The simulations are continued until 
$t = 100$s, which is much longer than the dynamical time scale in the inner region, where it is the shortest. 
Two spatial resolutions have been tried to see the numerical convergence: the normal resolution with 230 radial points and higher 
resolution with 460 points. Just as in Step 3, the grid width is determined by the scale height: 
$ \Delta r_{in} = 7.9 \times 10^{6} {\rm cm}$ and $ \Delta r_{out} = 2.0 \times 10^{8} {\rm cm}$ for the innermost and outermost grids, 
respectively, for the normal resolution, and they are twice finer for the high resolution case. The angular grid is uniform and has 60 
points in $0^\circ < \theta < 90^\circ$.

\begin{figure}
\vspace{15mm}
\epsscale{0.5}
\plotone{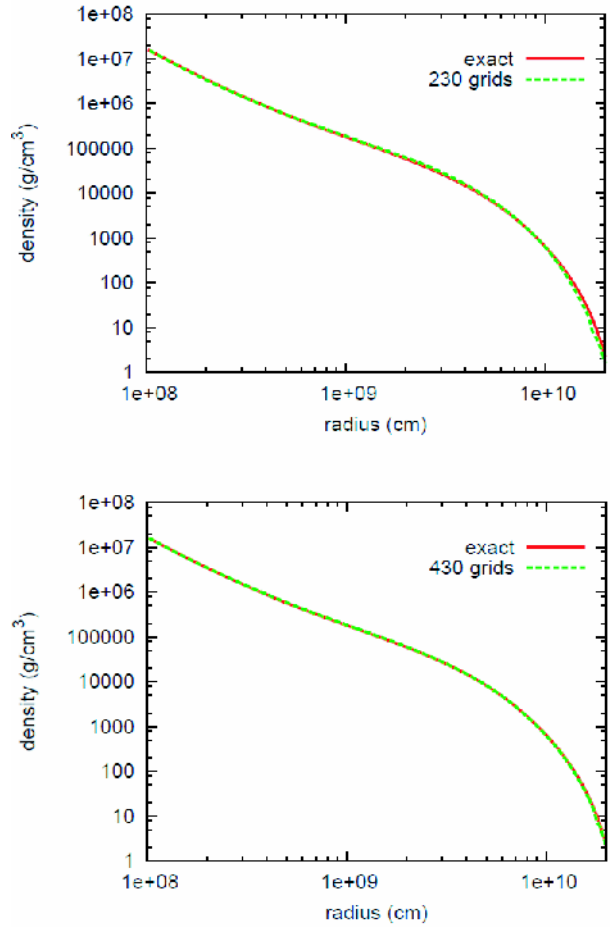}
\caption{The density profiles along the rotational axis at $t = 100$s after the long-term dynamical simulations of the envelope in
rotational equilibrium. Two spatial resolutions are employed with the upper panel showing the result for 230 radial grid points 
(green line) together with the initial profile (red line), whereas the middle panel corresponding to the result for 460 points. 
\label{f15}}
\end{figure}

Figure~\ref{f15} shows the density profiles along the rotational axis at the end of the simulations. The red lines are the 
profile obtained by the HSCF method and the green ones are numerical results. Deviations, which are inevitably induced by the mapping 
of the initial data as well as the difference in the finite-difference methods, are very small and it is indeed remarkable that 
the initial configuration is maintained in such a long time. This is clear evidence that both the hydrodynamics code and HSCF code are 
reliable. It should be also noted that these deviations are even smaller for the high resolution case (see the bottom panel of Figure~\ref{f15}).

\subsection{Sedov-Taylor Problems}\label{sedovtaylorpro}
In order to validate our AMR implementation, we have solved the Sedov-Taylor problem. Although our code is special relativistic, we use it 
in the non-relativistic regime here. It is also noted that a 2D grid is employed for the computation although the Sedov-Taylor problem is 
one dimensional, since our AMR code used in this paper is specialized for the jet simulation and based on the axisymmetric, two-dimensional grid.

In this simulation, the computational domain, $3 \times 10^8 {\rm cm} < r < 1.8 \times 10^{10} {\rm cm}$, is
covered by a uniform mesh with 100 radial grid points ($\Delta r = 1.8 \times 10^{8} {\rm cm}$) (this section is also adopted in cgs units). The internal energy of 
$E = 1.60 \times 10^{48} {\rm erg}$ is deposited initially in the central region of $r < 6.6 \times 10^{8} {\rm cm}$. 
The uniform density is set to be $\rho = 1 {\rm g/cm^3}$. We put a tiny specific internal energy 
($ \epsilon = 10^{-8} \times c^2 {\rm erg/g}$) uniformly except for the central region mentioned above just for numerical 
reasons. We adopt the $\gamma$-law EOS with $\gamma = 4/3$. We impose the free boundary condition both for the inner and outer boundaries. 
Although all $\theta$ derivatives vanish initially, they may evolve with time by numerical errors and we set 60 uniform angular grid points 
for $0^\circ \le \theta \le 90 ^\circ$. We impose the axisymmetry and equatorial symmetry on the z-axis and equatorial plane, respectively. 
We vary the resolution of the second level mesh to check the numerical convergence. The computations are terminated at 
$t = 20$s.

\begin{figure}
\vspace{15mm}
\epsscale{1.0}
\plotone{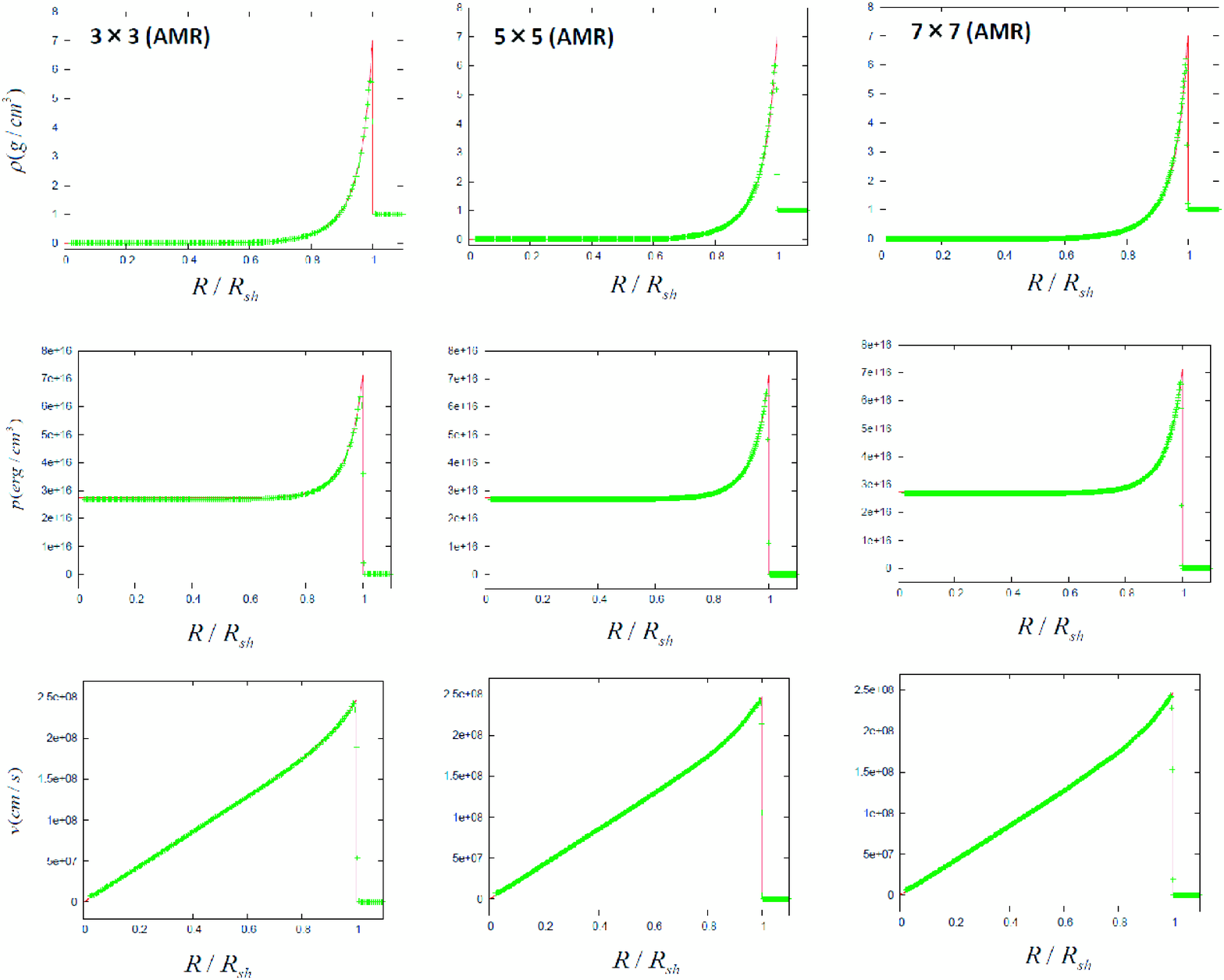}
\caption{The computed profiles of density (top), pressure (middle) and radial velocity (bottom) together with the exact solution (red lines)
for the Sedov-Taylor problem. The AMR resolution becomes higher from left to right.
\label{f16}}
\end{figure}

 Figure~\ref{f16} shows the numerical results (green dots) on the z-axis together with the analytical solutions (red lines). From 
left to right, different resolutions are employed for the second level mesh: 3 times, 5 times and 7 times finer in each coordinate
direction than the first level mesh for the left, middle and right columns, respectively. From top to bottom, the profiles of density, 
pressure and radial velocity on the z-axis are displayed, respectively. Note that the horizontal axis is a radius normalized by the 
radius of the shock front, $R_{sh} = 1.4 \times 10^{10} {\rm cm}$. As is evident in this figure, our AMR code successfully 
reproduces the analytical solution with an increasing sharpness of the discontinuity as the second level mesh becomes finer.

\subsection{A Pulse Propagating through Meshes of Different Refinement Levels}
In the AMR, there is a jump in resolution at the boundary between meshes of different refinement levels, which may produce unphysical waves. 
In order to make sure that the effect of the mesh boundary is negligible, we compute an pulse passing through the boundary adiabatically.

 The initial pulse profile is the same as that employed in \ref{subseconedimeisentflow}. Since the present AMR code is based on the
spherical grid, as mentioned earlier, we need in principle to reformulate the problem to accommodate a spherical wave. We can avoid
this issue , however, by the so-called thin shell approximation, in which the radial range of the computational domain is taken to be
much smaller than the radius itself and the coordinate curvature can be safely ignored. This convenient approximation has been widely 
used for plane-symmetric test problems on the spherical grid \citet[see e.g.][]{1997ApJ...475..720Y}. Here we take 
$r_{\star} = 10^{4}$ and $\Delta r_{\star} = 1.35$ for the representative radius and thickness of the shell, respectively. 
The initial density profile is given by $\rho_{0}(r-r_{\star})$ in Eq.~(\ref{isentroinirhopro}). We assume that all $\theta$ derivatives 
vanish initially. We employ a uniform angular mesh with 10 grid points in $0^\circ \le \theta \le 0.5 ^\circ$. We impose axisymmetry on 
the  z-axis and adopt the free boundary condition at $\theta = 0.5 ^\circ$. The radial extent of the computational domain is 
$-0.35+r_{\star} \le r \le 1+ r_{\star}$ and is covered by a uniform mesh with 210 radial grid points. For the inner region of 
$-0.35+r_{\star} \le r \le 0.3 + r_{\star}$, we deploy a second level mesh that is 3 times finer in each coordinate direction than 
the first level mesh.

\begin{figure}
\vspace{15mm}
\epsscale{0.5}
\plotone{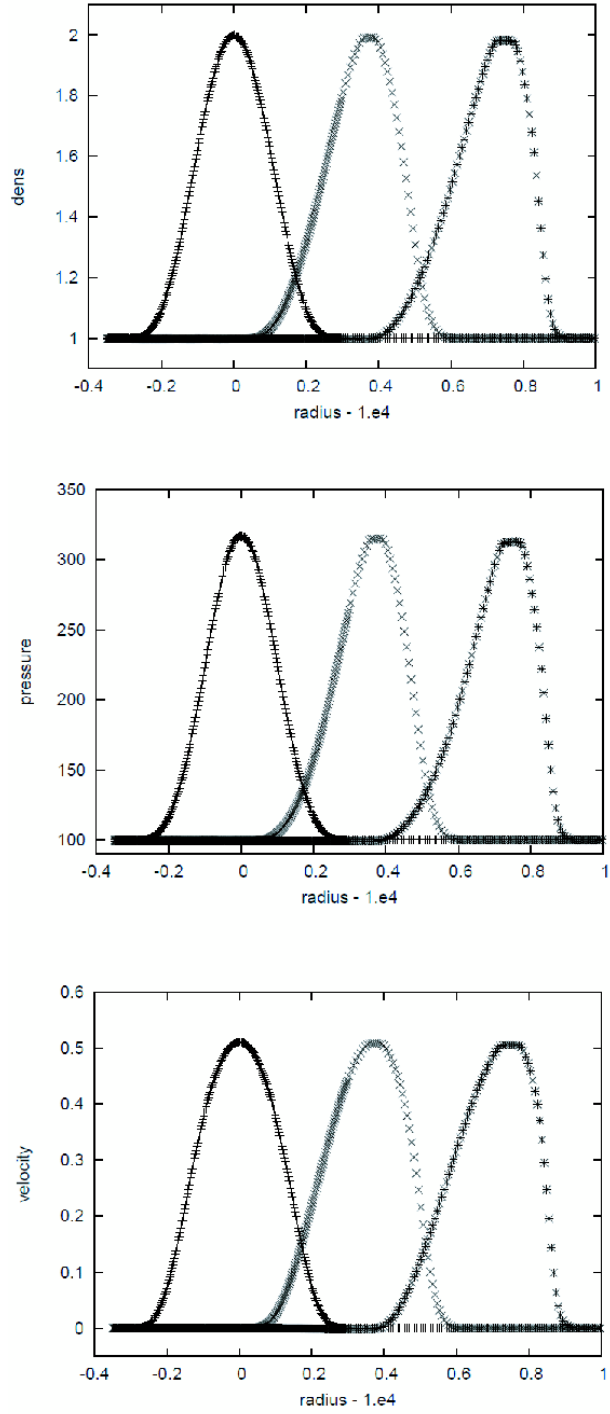}
\caption{The computed profiles of density (top), pressure (middle) and radial velocity (bottom) at $t = 0, 0.4, 0.8$ for the right-moving 
pulse. The mesh boundary is located at $ r - 10^{4} = 0.3$.
\label{f17}}
\end{figure}

Figure~\ref{f17} shows the numerical evolution of the pulse. No artificial waves are discernible when the pulse 
passes over the boundary ($r = 0.3 + r_{\star}$) between the meshes of different refinement levels (cf. Figure~\ref{f11}). 
Although the post-pulse state is not identical to the reference state, the difference (mainly caused by the grid curvature) is negligible 
(only $\sim 1 \%$).

In the above run, the pulse is initially located in the region covered by the mesh of higher refinement level and moves to the region 
of lower refinement level. We also run a simulation in the opposite case, i.e., the pulse is initially put in the region of 
lower refinement level (or the outer region) and moves to the inner region which is covered by the mesh of higher refinement level. 
This is realized by setting the initial density profile as $\rho_{0}(r-r_{\star}-0.7)$ in Eq.~(\ref{isentroinirhopro}) and determining 
the velocity distribution so that the Riemann invariant $J_{+} = {\rm const}$ in Eq.~(\ref{eqJpramin}) should be constant. 
The reference state and the amplitude and width of the pulse are unchanged. The numerical grids are also the same as above.

\begin{figure}
\vspace{15mm}
\epsscale{0.5}
\plotone{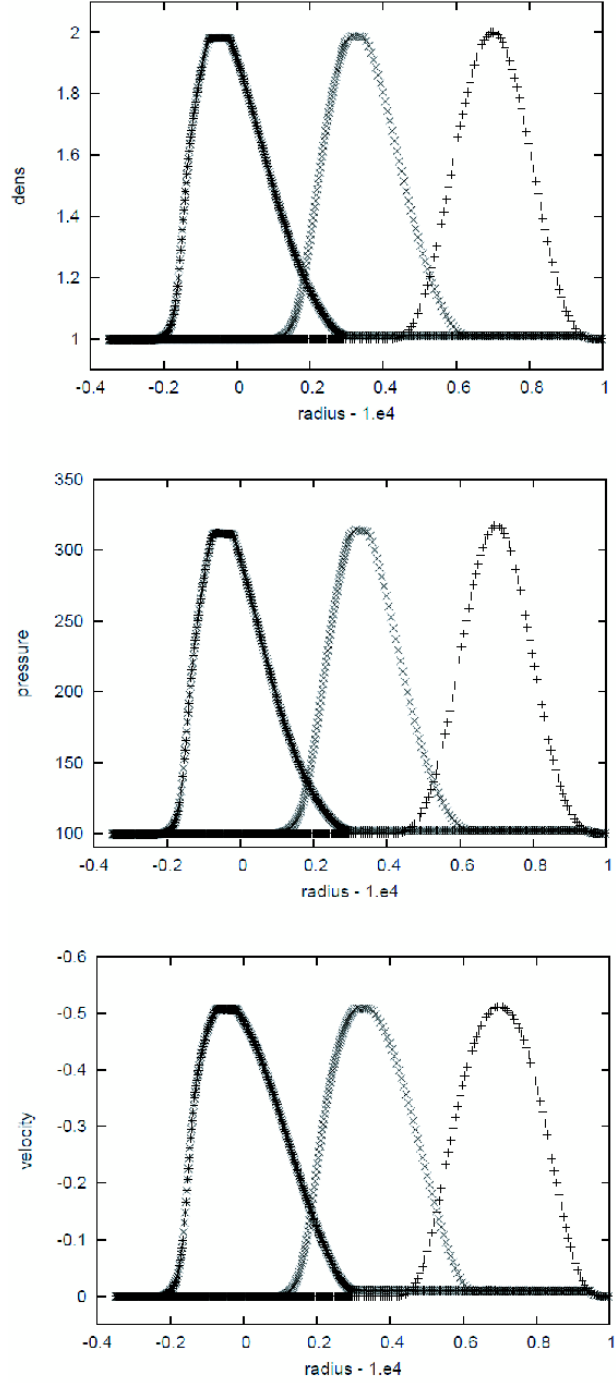}
\caption{The same as Figure~\ref{f17} but for the left-moving pulse. 
\label{f18}}
\end{figure}

 Figure~\ref{f18} shows the numerical results for the inward-moving pulse. Just as in the first case, the pulse passes 
through the mesh boundary, producing no discernible artificial wave. As expected, the pulse profile at the end of computation is a mirror
image of the one for the outward-moving pulse. These test results demonstrate the good behavior of our AMR code at the mesh boundary.

\subsection{Axisymmetric, Relativistic Jet Propagation in a Uniform Matter}
The last test is meant to investigate the effect of AMR resolution on relativistic jet propagations and has been frequently used in the 
literature. The computational domain covers the region $0.01 \le r \le 0.5$ and $0^\circ \le \theta \le 90^\circ$. Axisymmetry and equatorial 
symmetry are assumed and a uniform mesh with 150 radial grid points and 60 angular grid points are adopted as the first level mesh. 
We perform two simulations with different AMR resolutions: the second level meshes are 3 times and 9 times finer than the first level mesh, 
respectively. The relativistic jet is injected from the inner boundary by hand into the uniform medium. The injection parameters are 
$\rho_{0b} = 0.01$, $p_{b} = 1.70 \times 10^{-4}$ and $v_b = 0.99$. We employ $\gamma$-law EOS with $\gamma = 5/3$. The injection parameters 
give the Mach number of $M_b = 6$. The half opening angle of the jet is chosen as $\theta_{hop} = 9^\circ$. The reference state has a density 
and pressure of $\rho_{am} = 1.0$ and $p_{am} = 1.70 \times 10^{-4}$, respectively. Note that the ambient pressure is the same as the jet pressure. The simulation is terminated at $t = 2$.

\begin{figure}
\vspace{15mm}
\epsscale{1.0}
\plotone{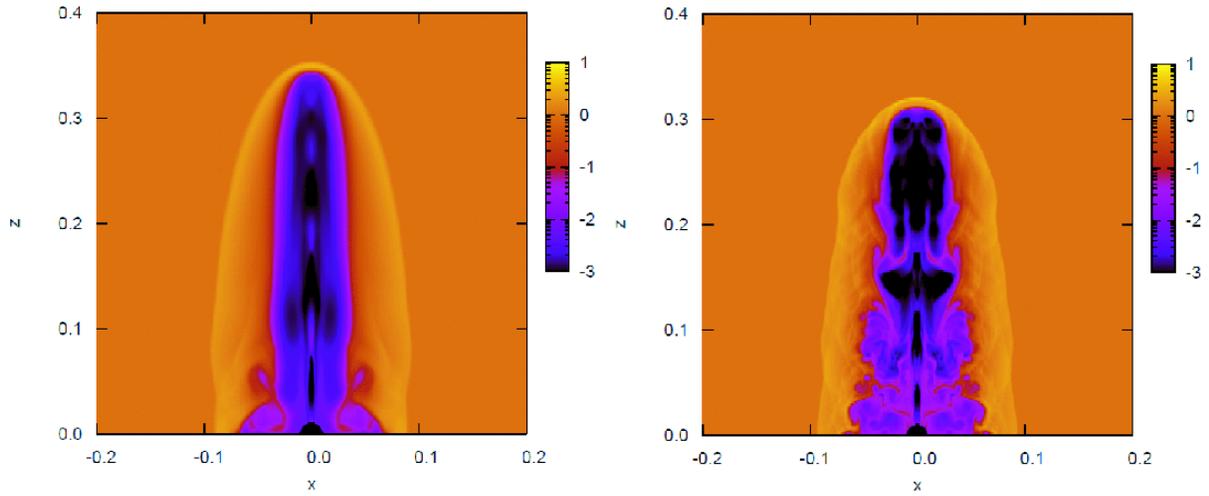}
\caption{Axisymmetric, relativistic jet propagations in a uniform medium. The density contours at $t = 2$ are displayed. 
The left panel shows the result for the case, in which the second level mesh is 3 times finer than the first level mesh and the 
right panel gives the result when a 9 times finer second level mesh is employed.
\label{f19}}
\end{figure}

Figure~\ref{f19} shows the density contours at the end of the computation, $t = 2$ for the two different AMR resolutions. The propagation 
of the forward shock wave and global structure are not very different between the two cases. It is evident that the higher resolution 
captures more complex internal structures.
 It is well known that the internal jet structure 
never converges as the numerical resolution gets better. This is due to the Kelvin-Helmholtz instability, for which the growth rate is greater for shorter wavelengths. Indeed the internal structures are very different between the two runs. The obtained numerical results in this test are qualitatively consistent with others in the literature (see e.g. Figure 11 in \citet{2006ApJS..164..255Z}).



\acknowledgments 






\begin{deluxetable}{cccc}
\tabletypesize{\scriptsize}
\rotate
\tablecaption{Numerical Errors for Different Resolutions in the 1D Isentropic-Flow Problem \label{tab1}} 
\tablewidth{0pt}
\startdata
\hline\hline
 Number of Grid Points   & $L^1$ Error (\%) & Convergence Rate ($\alpha$)\\
\hline
100 & 0.58 & - \\
200 & 0.17 & 1.83 \\
400 & 3.67E-2 & 2.17 \\
800 & 7.96E-3 & 2.20 \\
1600 & 1.80E-3 & 2.14 \\
3200 & 4.11E-4 & 2.13 \\
6400 & 9.60E-5 & 2.10 \\
\enddata
\tablecomments{The errors of density are evaluated at $t = 0.8$. In the right-most column, the powers in the expression 
$L^{1} \propto N^{- \alpha}$, where $N$ denotes the number of grids, are given.}
\end{deluxetable}
\begin{deluxetable}{cccc}
\tabletypesize{\scriptsize}
\rotate
\tablecaption{Numerical Errors for Different Resolutions in the 2D Isentropic-Flow Problem \label{tab2}} 
\tablewidth{0pt}
\startdata
\hline\hline\\
Number of Grid Points   & $\delta \rho_{0}$ (\%) & Convergence Rate ($\alpha$) \\
\hline
48 $\times$ 64  & 0.90 & - \\
96 $\times$ 128 & 0.24 & 1.93 \\
192 $\times$ 256 & 6.17E-2 & 1.93 \\
384 $\times$ 512 & 1.24E-2 & 2.32 \\
768 $\times$ 1024 & 2.70E-3  & 2.20\\
\enddata
\tablecomments{The errors of density are evaluated at $t = 2.4$. See the text for the definition of $\delta \rho_{0}$.
In the right-most column, the powers in the expression $ \delta \rho_{0} \propto N^{- \alpha}$, where $N$ denotes the number of grids, are given.}
\end{deluxetable}


\clearpage

\clearpage

\clearpage

\end{document}